\documentclass[]{aa}
\usepackage{epsfig}
\newcommand{\sub}[1]{_{\rm #1}}
\newcommand{\reference}[1]{}
\def\apj{{\em ApJ }}
\def\ApJ{{\em ApJ }}
\def\ApJS{{\em ApJS }}
\def\AandA{{\em A\&A }}
\def\aap{{\em A\&A }}
\def\MN{{\em MNRAS }}
\def\prl{{\em Phys.\ Rev.\ Lett.\ }}
\renewcommand{\S}{Sect. }
\newcommand{\Tab}{Table}
\newcommand{\changed}{}
\newcommand{\newchanged}{}
\newcommand{\thirdchanged}{}

\begin{document}

%

\title{Turbulent Velocity Structure in Molecular Clouds}

\author{V.~Ossenkopf\inst{1} \and{} M.-M.~Mac Low \inst{2}}

\institute{
1. Physikalisches Institut, Universit\"at zu K\"oln,
Z\"ulpicher Stra\ss{}e 77, D-50937 K\"oln, Germany
\and
Department of Astrophysics, American Museum of
Natural History, Central Park West at 79th Street, New York, NY,
10024-5192, USA
}

\date{Received: 15 December 2000; accepted: 22 April 2002}

\abstract{
We compare velocity structure observed in the Polaris Flare molecular
cloud at scales ranging from 0.015~pc to 20~pc to the velocity
structure of a suite of simulations of supersonic hydrodynamic and MHD
turbulence computed with the ZEUS MHD code. We examine different
methods of characterising the structure, including a scanning-beam
{\changed method that provides an objective measurement of Larson's} 
size-linewidth relation, structure functions, velocity and velocity
difference probability distribution functions (PDFs), and the
$\Delta$-variance wavelet transform, and use them to compare
models and observations.
\\
The $\Delta$-variance is most sensitive {\changed to} characteristic
scales and scaling laws, but is limited in {\changed its} 
application by {\changed a} lack of intensity weighting {\changed 
so that its results are easily dominated by observational noise
in maps with large empty areas. The scanning-beam
size-linewidth relation is more robust with respect to noisy data.
Obtaining the global velocity scaling behaviour requires that 
large-scale trends in the maps not be removed but treated as part of
the turbulent cascade.} We compare the true velocity PDF in our models
to simulated observations of velocity centroids and average line
profiles in optically thin lines, and find that the line profiles
reflect the true PDF better {\changed unless the map size is
comparable to the total line-of-sight thickness of the cloud.
Comparison of line profiles to velocity centroid PDFs can thus be used
to measure the line-of-sight depth of a cloud.}  
\\
The observed {\changed density and velocity} structure is consistent with
supersonic turbulence {\thirdchanged with a driving scale at or above the 
size of the molecular cloud and dissipative processes} 
below 0.05~pc. Ambipolar diffusion
could explain the dissipation. Over most of the observed range of scales the
velocity structure is that of a shock-dominated medium driven from
large scale. {\changed The velocity PDFs exclude small-scale driving
such as that from stellar outflows as a dominant process in the
observed region.  In the models, large-scale driving is the
only process that produces deviations from a Gaussian PDF shape
consistent with observations, almost independent of the strength of
driving or magnetic field.}  Strong
magnetic fields impose a clear anisotropy on the
velocity field, reducing the velocity variance in directions
perpendicular to the field.
}
\maketitle

\section{Introduction}

Attempts to characterise the physical state of molecular clouds by
comparison to simulations must rely on statistical
descriptions of the observations and the simulations.  In the last
decade, several techniques have been used to characterise
the observed radial velocity distribution in representative molecular
clouds, as reviewed by \cite{Goodman} and Miesch et al. (1999). 

First attempts to characterise the scaling behaviour of the velocity
structure started from one of the famous ``Larson's laws''
(\cite{Larson}), showing a power law relation between the size and the
linewidth measured for a molecular cloud. This relation has been
extended from integrated velocities to velocity fluctuations within
clouds by \cite{Miesch94}, providing similar power laws. On the other
hand hydrodynamic and magnetohydrodynamic (MHD) simulations have been
traditionally characterised using the probability distribution
functions (PDFs) of velocities and velocity differences
(e.g. \cite{Anselmet}).  Both approaches have been unified by
\cite{Lis} and \cite{Miesch99}, who measured the PDFs of line centroid
velocities and the scaling behaviour of PDFs of centroid velocity
differences as a function of lag for several star-forming clouds.

Mac Low \& Ossenkopf (2000, hereafter paper~I) used the
$\Delta$-variance, a multi-dimensional wavelet transform (\cite{Stutzki}), to
characterise both the density and the velocity structure of
interstellar turbulence simulations. For the density structure, a
direct comparison to the analysis of observed clouds provided by
Bensch et al. (2001a) was possible. In velocity space there is no
direct observational measure for the $\Delta$-variance. We need
additional tools for the quantification of the turbulent velocity
structure that can be determined with the same ease for observations
and simulations, and containing at least as much information
as the $\Delta$-variance.

In this paper we test five different methods on an observational data
set covering three steps of angular resolution in the Polaris Flare, a
translucent molecular cloud, and on a number of gas dynamical and MHD
simulations of interstellar turbulence.  The first two of the methods
characterise the total velocity distribution: the PDFs of the total
velocity distribution and of the line centroid velocities.  The other
three methods characterise the spatial distribution of velocities: a
generalised size-linewidth (``Larson'') relation, the dependence of
the low order moments of the centroid velocity difference PDF on lag,
and the $\Delta$-variance analysis.

Comparisons between observations and models have been made with a
simulation of mildly supersonic, decaying hydrodynamic turbulence
(\cite{Falgarone91,Falgarone94,f95}, Falgarone et al. 1995, Lis et al. 1996,
Lis et al. 1998, \cite{j98}, and \cite{pf00}), with MHD models of 
supersonic turbulence neglecting self-gravity by \cite{p98,p99} and 
\cite{p00}, with decaying and driven self-gravitating, hydrodynamic 
turbulence (Klessen 2000), and with various ad hoc models of 
turbulence (e.g.\ \cite{Dubinski}, Chappell \& Scalo 1999).
Observations suggest that supersonic, super-Alfv\'enic turbulence is an
appropriate physical model (see \cite{pn99}). Here, we
want to test whether it can reproduce the observed velocity structure
and find what we can learn about the physical conditions of turbulent
molecular clouds from comparison to a large set of models computed by
\cite{ml98} and \cite{ml99}.

In \S 2 we discuss the Polaris Flare observational data used 
here and the basic data processing applied. In \S 3 the different tools
for the analysis of the velocity structure are introduced and applied
to the observational data. The turbulence simulations used for comparison
are presented in \S 4 and the results of the velocity analysis
for these models are given in \S 5. \S 6 concludes with a discussion
on the physical state of the cloud based on our comparisons.

\section{Observational data}

Molecular line observations only determine line profiles, which give
the convolution of the radial velocity component with density along
the line of sight.  The situation becomes even more complicated if one
takes optical depth effects and spatially varying temperatures and
excitation levels into account. For the analysis provided here
we restrict ourselves to the assumption of constant excitation
conditions in an optically thin medium, so that the integrated line
intensity is a direct measure of the column density.

\subsection{Polaris Flare observations}

In Paper I, we compared our analysis of simulations with an analysis
of the intensity structure of multiscale observations of the Polaris
Flare performed by Bensch et al.\ (2001a).  As no prior analysis of the
velocity structure observed in this data has been done, we present
that here as one point of comparison to the simulation results.  

For the velocity field analysis we use three of the observational data
sets studied by Bensch et al. (2001a) in the analysis of the intensity
structure.  The Polaris Flare observations consist of a set of nested
maps obtained with the 1.2~m CfA telescope, the 3~m KOSMA, and the
30~m IRAM. The CfA data were taken in ${}^{12}$CO 1--0 at a spatial
resolution (HPBW) of 8.7' (Heithausen \& Thaddeus 1990); the KOSMA
observations used the ${}^{12}$CO 2--1 transition at 2.2' resolution
(Bensch et al. 2001a); and IRAM observations of the MCLD~123.5+24.9
region in the Polaris Flare were taken in the two lower transitions of
${}^{12}$CO and ${}^{13}$CO and in the 1--0 transition of C${}^{18}$O
within the IRAM key-project ``Small-scale structure of
pre-star-forming regions'' (Falgarone et al. 1998). To discuss a
consistent set of observations for all maps we restrict ourselves to
the ${}^{12}$CO IRAM data. Because of the higher signal-to-noise
ratio, we only use the 1--0 lines taken at 0.35' resolution.  Assuming
a distance to the Polaris Flare cloud of 150~pc\footnote{
\cite{Heithausen90} estimated a distance to the Polaris Flare cloud of
240pc whereas \cite{Zagury} derived 105--125pc.}, the telescope
resolutions translate into physical resolutions of 0.38~pc, 0.09~pc,
and 0.015~pc respectively.  Altogether these observations provide a
data set covering more than three decades in linear resolution -- from
0.015~pc to about 50~pc.

A major problem when combining these data is the different channel
width, noise, and baseline behaviour of the different instruments
and observational runs. The
IRAM observations show an rms noise of 0.5~K at 0.05~km/s, with some
large scale trends in the noise indicating that either the spectral or
the spatial baseline is not optimal. The KOSMA data have an rms of
0.4~K at 0.05~km/s throughout the whole map. The CfA data show an rms
noise of 0.1~K at 0.65~km/s, and show both a variation of the
absolute noise level throughout the map, and some areas with slightly
negative intensities, indicating imperfect baselines. This opens up
some uncertainties when combining noise sensitive results from the
three data cubes.

\subsection{Line windowing}

The basic problem in the deduction of the velocity structure is the
finite signal-to-noise ratio in each observed map, often combined with
an imperfect spectral baseline.  Due to small slopes and variations of
the baselines, and a slightly variable noise throughout the spectrum,
the determination of the velocity structure from the spectra is
sensitive to the exact selection of the spectral window around the
line considered. The influence of this effect increases from the line
centroid velocities to the higher moments like the variances or the
kurtosis.  A detailed discussion of these problems in the
determination of the centroid velocity was provided by \cite{Miesch94}
and \cite{Miesch95}.

We have tested three different methods for the selection of that part
of the spectrum containing as much information as possible about the
line but least influenced by baseline uncertainties. First we applied
a global windowing technique defining a minimum velocity range
covering all noticeable emission.  We used this global windowing as a
first rough constraint to the velocity space, including all channels
where eye inspection might still guess some line contribution.  A
second method that we have extensively tested was to search for the
line contribution interval by using the first zeros at the flanks of
the lines. We found, however, that too many lines are either
relatively weak or break up into several components so that zeros
occur within the line.  Even the centroid velocities could not be
reliably determined from this approach.  As a third approach, we have
used a typical criterion for noticeable emission, such as emission
above a 3 $\sigma$ noise level. It turned out, however, that for most
values of the significance level, we clearly miss part of the
information from weak but broad lines in thin outer regions that
inspection by eye would still count as a part of the line.

Thus we show for all our observational results large
error bars given by two extremes of this criterion: we
count either all contributions within the global spectral window, or only
channels above the maximum noise level given by the largest negative
value.  As representative intermediate values we show in all plots the
results using all positive contributions above 1$\sigma$ of the noise.
This third level provides a reasonable intermediate value and
the parameters from the data analysis below show only a small
variation when changing the noise cut level around this value. One
should however keep in mind that we do not know the best treatment
of the noise so that this line should be considered only as guiding
the eye but not as the best representative or average value.
\label{sect_windowing}

\subsection{Large-scale trends}

\mbox{}\cite{Miesch94} extensively discussed the removal of large scale trends
in the centroid velocity maps to get a significant description of the
turbulent velocity fluctuations undisturbed by any large systematic
motions. We will not follow their method here when dealing
with the Polaris Flare data. Due to the nested nature of the different
maps, any large scale motion on a smaller map is only a velocity fluctuation
on the larger map. As \cite{Bensch} have shown for the intensity maps,
a smooth transition in the scaling behaviour of the three maps is only
possible if large scale trends are not removed. 

Furthermore, there is no clear separation between turbulent and systematic 
motions. Any assumed separation scale is arbitrary as long as there is
no physical process such as energy injection at that scale.  Removal
of velocities at certain 
scales might prevent understanding of the underlying processes.
Even the largest systematic motions, like Galactic rotation, may be part
of the turbulent cascade if they inject energy into the system.

It could be justifiable to remove the large scale trends for the
star-forming regions considered by \cite{Miesch94} if the turbulence
there were only driven by small-scale star-formation activity, and not
by large-scale motions.  However, that is not proven even there, and
we would certainly miss the main physics by applying the same kind of
separation for the Polaris Flare data, where the turbulence probably
is driven by motions on the largest scales (see discussion in Paper
I).

\section{Statistical descriptions of velocity structure and their applicability
to observations}

\subsection{Size-linewidth relation}

A traditional measure for the spatial velocity distribution is the
size-linewidth relation for clouds identified by Larson
(1981) and obtained by many observers since then.  To measure this,
one has to define objects within a map, such as molecular clouds or
clumps within clouds, and relate the effective linewidths measured for
these objects to their characteristic sizes.  This method has been
used to study a large variety of clouds, clumps, and cores
(e.g. \cite{myers}, \cite{caselli}, \cite{peng}).  A comprehensive
recent overview including a careful estimate of many possible errors
was given by Goodman et al.\ (1998).  Most studies obtain power laws
\begin{equation}
\Delta v_{\rm obs} \propto R^\gamma \quad {\rm with} \; \gamma=0.2 \dots 0.7
\end{equation}
over wide spatial ranges, where $R$ denotes the effective radius of
the object.  A search for Larson-type relations in turbulence
simulations was performed by \cite{Semadeni}.

A major problem in the computation of these size-linewidth relations
is the somewhat arbitrary definition of the objects in the observed
position-velocity space that are considered to give definite values for
sizes and linewidths.  This definition is obvious, though dynamically
arbitrary, for isolated molecular clouds, but difficult when selecting
clumps within a cloud, and completely impractical for 
filamentary, turbulent cloud structures.

As an alternative, one can compute a size-linewidth relation by
measuring the average linewidths within telescope beams of varying
size -- effectively the intensity-weighted velocity dispersion within
a varying radius.  Practically, the observed map is scanned with a
Gaussian of varying size and the average linewidth is determined for
each size. In averaging, each position is weighted by its total
intensity so that the lower significance of ``empty'' regions is taken
into account.  This method can be easily applied to each data set and
the results are directly comparable to the traditional size-linewidth
relations.

We face, however, the problem that the velocity dispersion for each
virtual beam depends not only on its size, but also on the cloud depth along
the line of sight.  This depth enters by changing the local linewidth at
each point. To separate the two length scales we have applied the analysis
both to the total velocity dispersion within the Gaussian, and to the
dispersion of the line centroids observed at each point in the
map. In the latter case, we hope to remove the influence of the cloud 
depth.

\begin{figure}
\centering
\epsfig{file=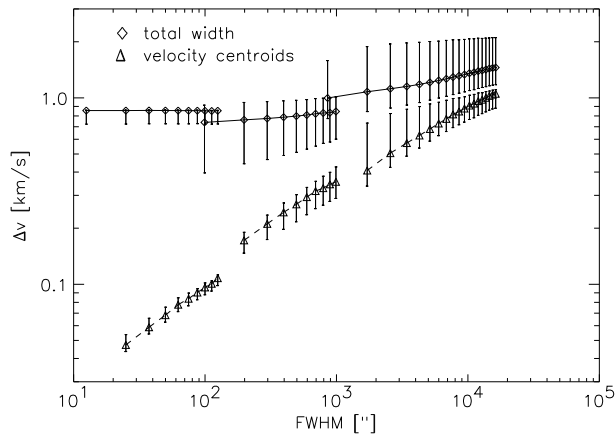,angle=90,width=\columnwidth}
\caption{Size-linewidth relation for the 
Polaris Flare CO observations with IRAM (smallest scales), 
KOSMA, and the CfA 1.2m telescope (largest scale,
see \cite{Bensch} for details). The diamonds show the relation when 
the linewidth for a given scale is integrated from the
total linewidths at each point. The triangles represent the
widths when only the dispersions of the centroids of the lines are measured.
{\newchanged The error bars do not represent true local errors but
the two extreme cases of the line windowing as discussed in
\S \ref{sect_windowing}.}}
\label{polaris_larson}
\end{figure}

In Fig. \ref{polaris_larson} we show both relations for the
Polaris Flare observations.  The two extreme approaches for the noise
treatment (taking all emission from the spectral window or only
emission above the maximum noise level) produce relatively large error
bars in the plots for both the total velocity dispersion and the
centroid velocity dispersion.  The CfA data with their low spectral
resolution in particular have error bars of up to a factor of two.
Within the errors, however, we find a unique smooth behaviour in both
quantities from the smallest to the largest scales.  The KOSMA data do
deviate somewhat from the other two sets. This can be partially
attributed to the different CO transition observed by that
telescope. We have checked this effect by computing the same plot for
the IRAM ${}^{12}$CO 2--1 data. Although the results were shifted
relative to the IRAM 1--0 data in the same direction as the KOSMA
data, they do not line up exactly with the KOSMA result. Thus the shift
is probably also influenced by the different noise behaviour.

For the size-linewidth relation based on the velocity centroids, we
find one power law stretching over three orders of magnitude connecting
the three different maps. The average velocity variances range from below
the thermal linewidth up to about 1~km~s$^{-1}$. The common slope is
given by $\gamma=0.50\pm 0.04$. However, the data are also consistent with a
reduction of the slope down to 0.24 at the largest scales, if the full
extent of the error bars is taken into account.

In the size-linewidth relationship integrated from 
the full local linewidths, there is a transition of the slope from
almost zero at scales below 10' to 0.2 at the full size of the
flare. The plot shows that the total linewidths are dominated by the
line-of-sight integration up to the largest scales.  

Although the slopes measured with this method are very shallow, they
do appear to show the change of slope interpreted by \cite{Goodman} as
a transition to coherent behaviour below about 0.5 pc.  As 
the findings of Goodman et al.\ are also based on the total linewidths
this suggests that the change might rather reflect the
transition from a regime where single separated clumps are identified,
to measurements of a superposition of substructures at smaller scales.
\label{sect_larson}

\subsection{Velocity probability distribution function}

\label{subsect-vpdf}
Another quantity characterising the velocity structure both in
observational data and in turbulence simulations is the probability
distribution function (PDF) of velocities. Although it contains no
information on the spatial correlation in velocity space like the
size-linewidth relation or the $\Delta$-variance, it shows
complementary properties, like the degree of intermittency in the
turbulent structure (\cite{Falgarone90}). The shape of the wings of
the velocity PDF is thought to be diagnostic of intermittency, 
where the increasing degrees of intermittency produces a transition
from Gaussian to exponential wings.
Two-dimensional Burgers turbulence simulations by
\cite{Chappell}, neglecting pressure forces, showed Gaussian
velocity PDFs for models of decaying turbulence and exponential wings
for models driven by strong stellar winds. 

Due to the limited amount of information available from molecular
lines, there is no direct way to deduce the velocity PDF from
observations.  One approach to deducing the velocity PDFs is
computation of the distribution of line centroid velocities (Kleiner
\& Dickmann 1985; Miesch \& Bally 1994; Miesch et al.\ 1999).  This
method can also include some information on spatial correlation as
discussed in \S~\ref{sect_diffpdfs}.  However, the higher moments of
the centroid PDF are very sensitive to the observational restrictions
discussed in \S~\ref{sect_windowing}.

Another method was introduced by Falgarone \& Phillips (1990), who
estimated velocity PDFs from high signal-to-noise observations of
single line profiles. Investigating the statistical moments of
profiles, Falgarone et al.\ (1994) found no simple Gaussian behaviour
for many observations and provided a first comparison with three-dimensional
(3D) hydrodynamic simulations. Most of their PDFs could be represented by a
superposition of two Gaussians where the wing component had about
three times the width of the core component.  Unfortunately, their
method is only reliable for optically thin transitions at a very high
signal to noise.  We test both methods here, starting with the
centroid velocity PDF.

\subsubsection{Centroid velocity PDFs}

In computing the centroid velocity PDF for a map one can either assign
the same weight to each point in the map, or weight the different
contributions by the intensities measured at that point. 
{\changed We find that the PDFs retain similar shape and the same
wing behaviour with both methods, and} therefore 
use intensity weighting
in the following analysis, as it is less influenced by observational
noise.  We have also used normal histograms here, instead of the more
sophisticated Johnson PDF estimator applied by \cite{Miesch99} because
the error bars present from the uncertainty about the noise treatment
greatly exceed the influence of the numerical PDF estimator. 

\begin{figure}
\centering
\epsfig{file=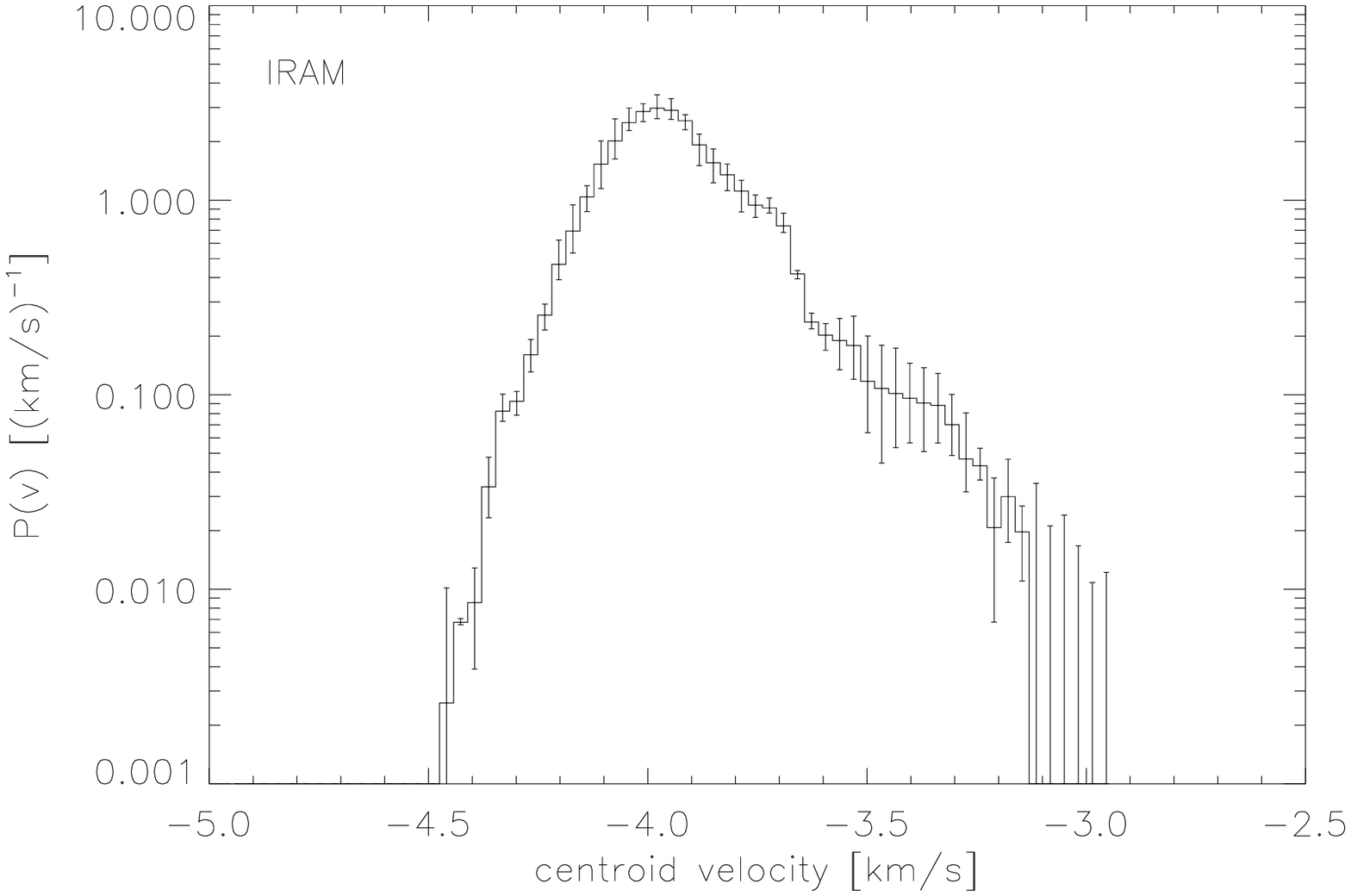,angle=90,width=\columnwidth}
\epsfig{file=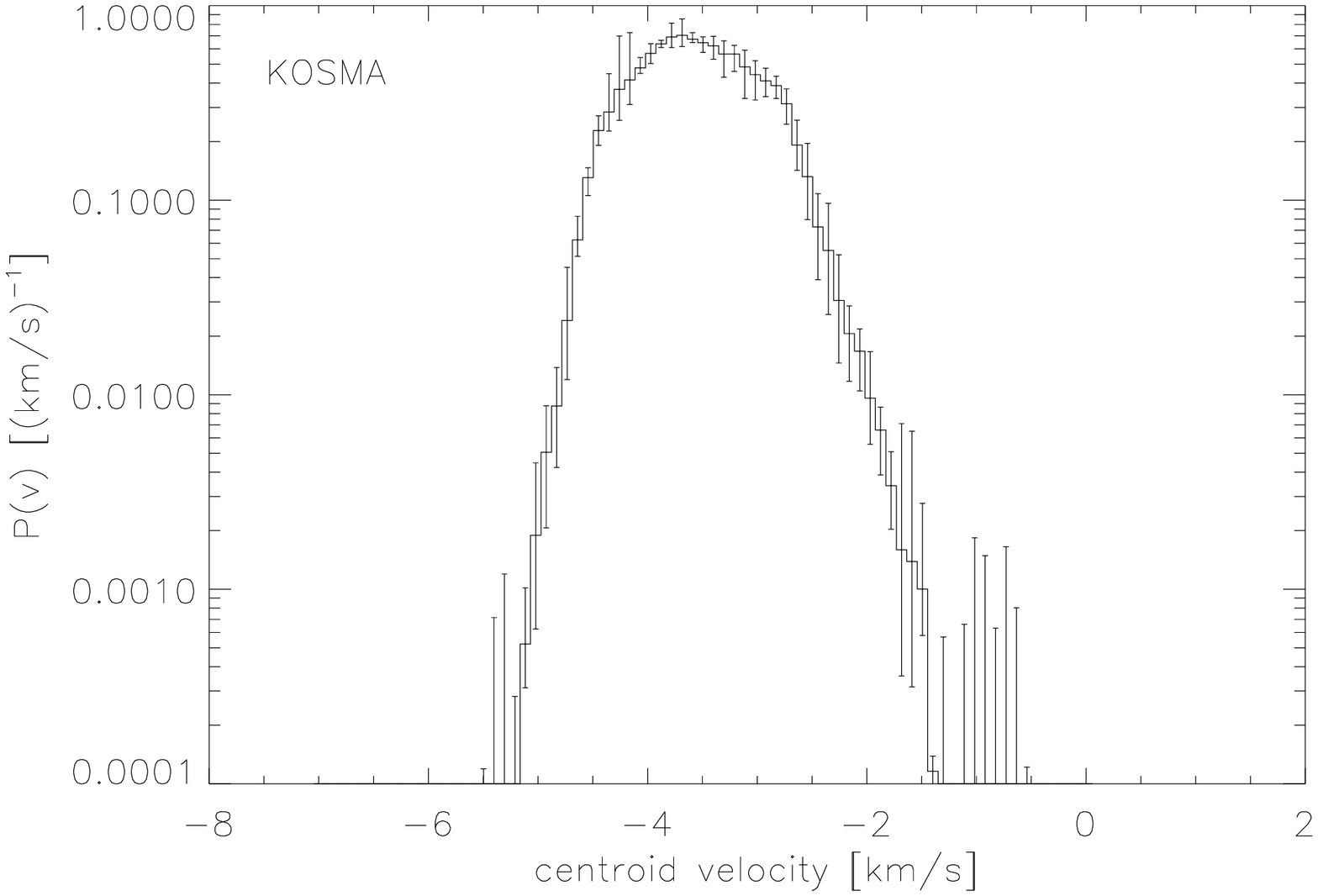,angle=90,width=\columnwidth}
\epsfig{file=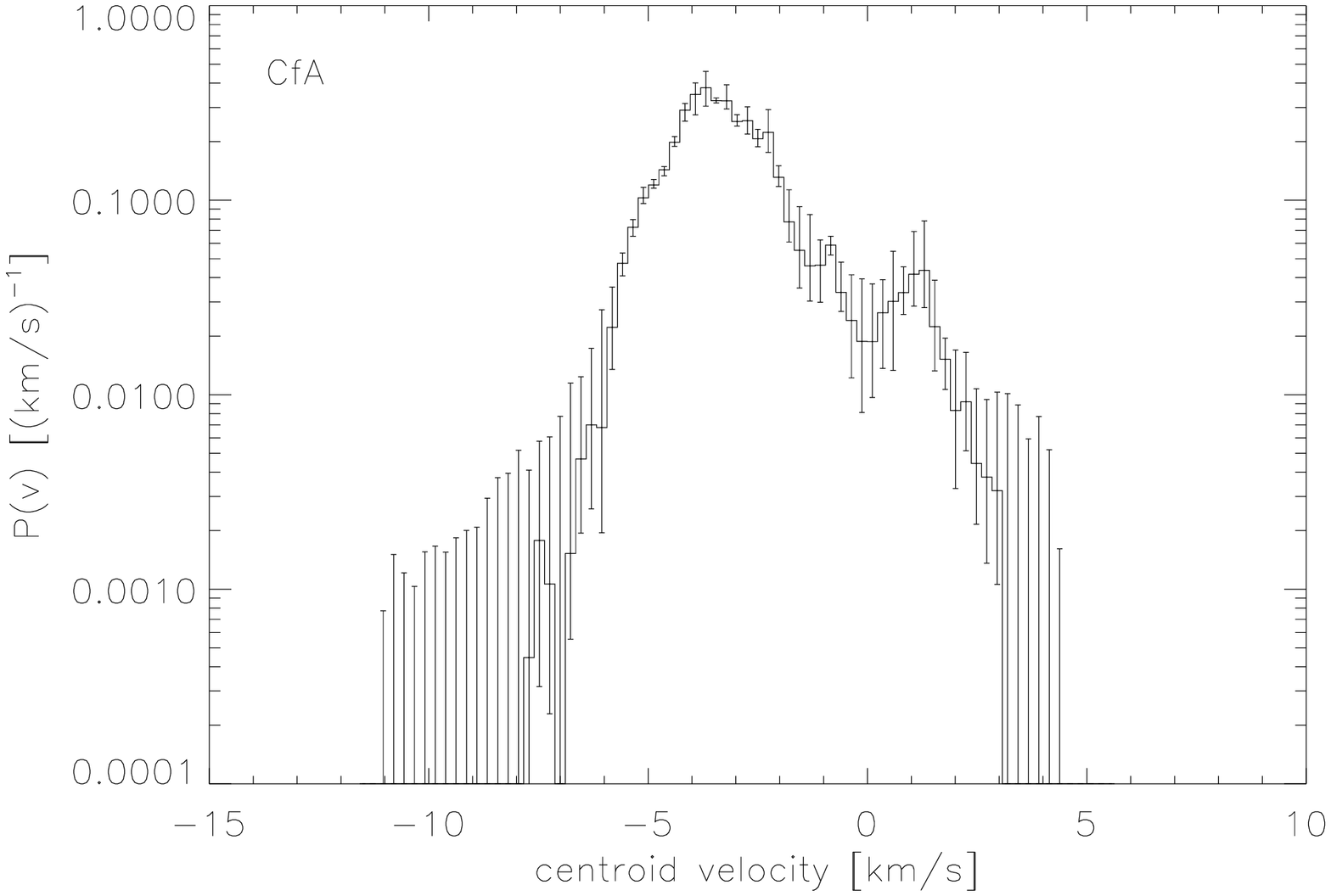,angle=90,width=\columnwidth}
\caption{Probability density distribution of centroid velocities for
the three Polaris Flare data cubes. The error bars show the
deviation introduced by different treatments of the observational
noise, as discussed in the text.}
\label{fig_pdfs}
\end{figure}

\begin{table*}
\caption{Parameters of the centroid velocity PDFs and the average
line profiles in the three Polaris Flare maps
\label{tab_polarismoments}}
\begin{tabular}[h]{llllll}
PDF type & telescope & std. deviation & kurtosis & core width$^{\mathrm a}$ & wing width$^{\mathrm a}$\\
&& [km/s] & & [km/s] & [km/s]\\
\hline
centroid PDFs & IRAM & $0.17 \pm 0.01$ & $5.1 \pm 0.2$ & $0.130 \pm 0.005$ & $0.24 \pm 0.05$\\
& KOSMA & $0.53 \pm 0.03$ & $2.7 \pm 0.1$ & $ 0.57 \pm 0.05$ & $0.41 \pm 0.03$\\
& CfA & $1.6 \pm 0.2$ & $5.0 \pm 0.3$ & $ 1.02 \pm 0.05$ & $2.2 \pm 0.5$\\ 
line profiles & IRAM & 0.88 & 2.4 & $0.98 \pm 0.04$ & $0.80 \pm 0.07$\\
& KOSMA & 1.06 & 2.7 & $1.10 \pm 0.01$ & $1.03 \pm 0.05$ \\
& CfA & 2.3 & 3.8 & $1.8 \pm 0.1$ & $2.7 \pm 0.3$\\
\hline
\end{tabular}
\\ \footnotesize
${}^{\mathrm a}$ standard deviation of the Gaussian fit
\normalsize
\end{table*}

Fig.~\ref{fig_pdfs} shows the centroid velocity PDFs for the three
data sets.  We find that the IRAM and the CfA data are characterised
by an asymmetry of the velocity distribution, indicating some kind of
large-scale flow within the mapped region. Looking at the wings of the
distributions, however, all three data sets are consistent with a
Gaussian, which would appear as a parabola in the lin-log plots shown.
Only at the scale of the CfA map is a definite conclusion not
possible, due to the large error bars.

Beyond this phenomenological approach, the shape of the PDFs can be 
quantified by their statistical moments. The {\changed most frequently
used moments are
\begin{eqnarray}
\langle v\sub{c} \rangle &=& \int_{-\infty}^{\infty} dv\sub{c}
P(v\sub{c}) v\sub{c}\\ 
\sigma^2 &=& \int_{-\infty}^{\infty} dv\sub{c} P(v\sub{c})
[v\sub{c}-\langle v\sub{c} \rangle]^2\\ 
K &=& {1 \over \sigma^4} \int_{-\infty}^{\infty} dv\sub{c} P(v\sub{c})
[v\sub{c}-\langle v\sub{c} \rangle]^4 
\end{eqnarray}
where $\langle v\sub{c} \rangle$ is the mean, $\sigma^2$ the variance,
and $K$ the kurtosis of the distribution.  The
probability distribution function $P$ is normalised to unity.
The variance is a measure for the total
turbulent mixing energy, while the kurtosis characterises the
deviation from a Gaussian profile.  It takes a value of three for a
Gaussian distribution, and six for a distribution with exponential
wings. Values between about 2.7 and 3.0 are still consistent with 
a Gaussian which is truncated due to a finite sample size. Shallow
wings become obvious for kurtosis values above about 3.4.}

We restrict the analysis to {\changed these low moments
because 
higher moments become increasingly uncertain} due to the influence of
observational noise, non-perfect spectral baselines, and error-beam
pickup (see e.g. Bensch et al.\ 2001b). Without extremely high signal
to noise data it is impossible to obtain reliable constraints on the
spatial variation even for the next higher velocity moments.
{\changed The situation is even worse for}
methods like the spectral correlation function 
(\cite{Rosolowsky}) that include all details of the line profiles. 

In \Tab~\ref{tab_polarismoments} we give the standard deviation (square
root of the variance) and kurtosis for the three PDFs. Only the 
KOSMA map shows an almost Gaussian distribution of velocity centroids
with slightly steeper-than-Gaussian wings. The kurtosis for the other two maps
is clearly larger than Gaussian. We have tested whether we can 
reproduce this by the superposition of two Gaussians as proposed by
Falgarone et al. (1991). Within the error bars we always obtained good fits.
All wings can be reproduced by a Gaussian. The widths of the core and 
the wing component obtained from this fit are also given in
Table~\ref{tab_polarismoments}. In contrast to the width ratio of about three
obtained by Falgarone et al. we find a ratio of about two.

Only a few of the observed maps analysed by \cite{Miesch99} showed 
approximately Gaussian centroid velocity PDFs, while the majority had
PDFs with shallower wings that could be fitted with either exponential
laws or power laws. Fitting the exponent, \cite{Miesch99} obtained two
different values when treating either the whole distribution or
only the wings. This is similar to our fit of the observational
data with two different Gaussians for the wing and the core component.

In the Polaris Flare maps the wings of all distributions can be
represented by Gaussians. However, the total distributions sometimes
deviate considerably from Gaussian behaviour resulting in a Gaussian
kurtosis value at intermediate scales (the KOSMA map) compared to
significantly larger kurtosis at smaller and larger scales. Besides the
effect of observational errors, this might represent the influence of
systematic velocity trends across the mapped region which appear
mainly in the core component but not in the wing. We don't expect
these trends in the isotropic simulations analysed in \S~\ref{sect_simpdfs}.
\label{sect_centpdfs}

\subsubsection{Average line profiles}

\begin{figure}
\centering
\epsfig{file=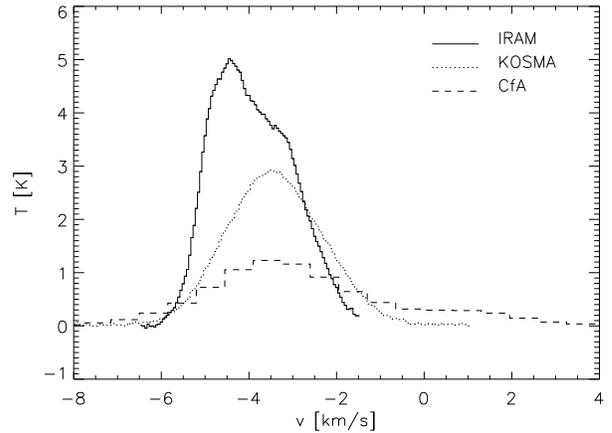,angle=90,width=\columnwidth}
\caption{Average line profiles for the three Polaris Flare data
cubes. The solid line shows the IRAM data, the dotted line the
average profile in the KOSMA map multiplied by 2 and the dashed line
the CfA data multiplied by 8.}
\label{fig_avlines}
\end{figure}

In Fig. \ref{fig_avlines} we show the average line profiles for all
three data sets. The corresponding moments and fit parameters of the
distributions are also given in \Tab~\ref{tab_polarismoments}. The
average line profile automatically contains the weighting of each
velocity contribution by its intensity as discussed above for
the centroid PDFs, so that we can compare both. 

{\changed The general shape is similar but the
line profiles are much broader and the peak positions are
not at the same velocity as in the PDFs.} 
\begin{figure}
\centering
\epsfig{file=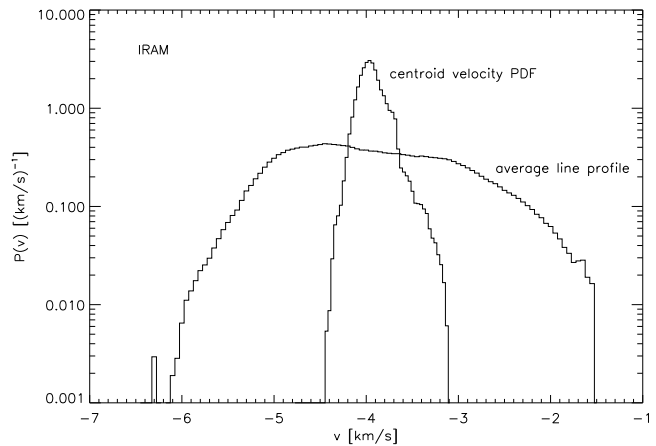,angle=90,width=\columnwidth}
\caption{Comparison of the total velocity PDF given by the
average line profile and the centroid velocity PDF for the
IRAM data.}
\label{fig_irampdfs}
\end{figure}
For a detailed comparison, we plot both the average line profile and
the centroid velocity distribution for the IRAM data on the same
logarithmic scale in Fig.\ \ref{fig_irampdfs}.  
The ratio
between average line profile width and centroid velocity PDF width
measured from the Gaussian fitted to the wings of the
distributions is about 3.4 for the IRAM map, 2.2 for the KOSMA map,
and 1.5 for the CfA map. Using the variance of the full distributions
we get ratios of 5.2, 2.0, and 1.4 respectively for the three maps.
The difference of the two ratios for the IRAM map corresponds to some 
large scale velocity flow on that scale producing the irregular PDF 
core seen in Fig.\ \ref{fig_irampdfs}.

The variation of this ratio with the size of the map is again naturally 
explained by the two different length scales involved: the 
line-of-sight integration and the size of the map.  In the IRAM data,
the small size of the map provides a relatively narrow
centroid velocity PDF compared to the broad average line profile
determined by the line-of-sight integration through the full depth of
the cloud. In contrast, the thickness of the cloud will certainly be
smaller than the full extent of the CfA map. This is also indicated 
by the approaching slopes of the two different variances
within beams of varying size in Fig.\ \ref{polaris_larson}
at large scales. To compare the ratios obtained here with
turbulence simulations in model cubes we have to consider 
scales where the map size is about equal to the thickness of the cloud.
From the intensity maps of the clouds and Fig.\ \ref{polaris_larson}
we estimate a thickness corresponding to about $2^\circ$ in angular
scale. The resulting typical value for the width ratio that we should 
reproduce
in the turbulence simulations then falls between about 1.5 and 1.6.
We will see that several but not all turbulence models show
such values.

We can resolve the long lasting dispute over whether to use the
average line profiles or the centroid velocity PDF as a measure for
the 3D velocity PDF. The answer is determined by the
size scales involved in the observations. Since the velocity centroids
ignore the integration along the line of sight, they provide the
correct distribution 
{\changed when} the map size is larger {\changed than or}
comparable to the thickness of the cloud, while for small maps, the line
profiles provide the better average, because they include a larger
sample from the {\changed longer} line-of-sight integration.

We have to mention 
{\changed two caveats. First, optical depth effects can broaden lines.}
However,
\cite{Bensch} have found that the ${}^{13}$CO and ${}^{12}$CO Polaris
data show the same spatial scaling laws, although the maps
differ, {\changed suggesting}
that optical depth effects play
only a minor role.  
{\changed Second, the two methods can only be equivalent for an isotropic
medium, which is not guaranteed.}
\label{sect_pdfs}

\subsection{Velocity difference PDFs}

From the velocity centroid maps one can also extract information on
the distribution of scales in the velocity field by considering PDFs
of velocity differences between points separated by different lags
(distances).  This provides independent information on the structure
of the velocity field.  Investigation of the PDF of velocity
differences as a function of spatial separation has been pursued by
Miesch \& Scalo (1995), Lis et al.\ (1998), and Miesch et al.\
(1999). For a discussion of the details and the application to several
molecular clouds we refer to Miesch et al.\ (1999).

Here we don't study the full PDF of centroid velocity differences but
the variation of the first statistical moments of this PDF as a
function of lag between the two points considered.  Because of the
symmetry of the velocity differences, all odd moments vanish.  The
first two non-zero moments of the velocity difference distribution are
the variance and the kurtosis:
\begin{eqnarray}
\sigma^2(L)&=&{\displaystyle \int_{\rm map} \hspace*{-0.3cm}d^2\vec{r} 
\int_{|\vec{r}-\vec{r'}|=L} \hspace*{-0.8cm}d^2\vec{r'}\,
f(\vec{r}) f(\vec{r'}) \left[v\sub{c}(\vec{r})-v\sub{c}(\vec{r'})\right]^2
\over \displaystyle \int_{\rm map} \hspace*{-0.3cm}d^2\vec{r} 
\int_{|\vec{r}-\vec{r'}|=L} \hspace*{-0.8cm}d^2\vec{r'}\,
f(\vec{r}) f(\vec{r'}) } \label{eq_struct} \\
K(L)&=&{\displaystyle \int_{\rm map} \hspace*{-0.3cm}d^2\vec{r} 
\int_{|\vec{r}-\vec{r'}|=L} \hspace*{-0.8cm}d^2\vec{r'}\,
f(\vec{r}) f(\vec{r'}) \left[v\sub{c}(\vec{r})-v\sub{c}(\vec{r'})\right]^4
\over \displaystyle \sigma^4(L) \,\int_{\rm map} \hspace*{-0.3cm}d^2\vec{r} 
\int_{|\vec{r}-\vec{r'}|=L} \hspace*{-0.8cm}d^2\vec{r'}\,
f(\vec{r}) f(\vec{r'}) }.
\end{eqnarray}
Integrations over the spatial vectors $\vec{r}$ and $\vec{r'}$  scan the whole map.
The contribution of the velocity difference between the points
$\vec{r}$ and $\vec{r'}$ is weighted by weighting factors $f$. 
{\changed We compared equal weighting as used by Miesch et al.\
(1999), weighting by the geometric mean, and by the product of the two
intensities, and find little difference.}  In the following we use the
weighting by the geometric mean, as it is a linear intensity weighting
for each term, as in the case of the PDFs discussed above.

\begin{figure}
\centering
\epsfig{file=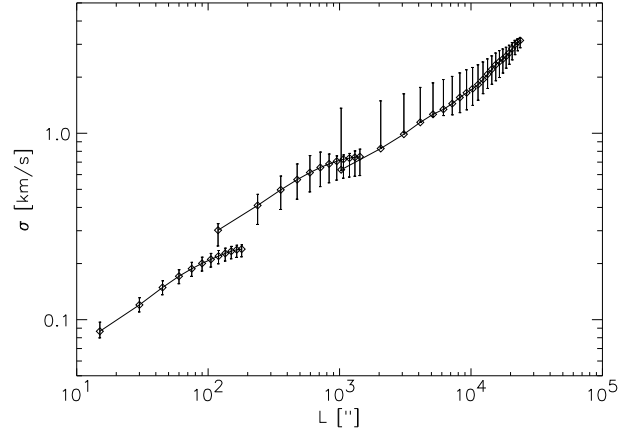,angle=90,width=\columnwidth}
\caption{The standard deviation of the velocity difference PDF as
a function of lag.}
\label{fig_variance}
\end{figure}

The quantity given by the variance of the two-point PDF $\sigma^2(L)$
in Eq.\ (\ref{eq_struct}) as a function of the lag between the points
$L$ is identical to the ordinary structure function as used e.g. by
\cite{Miesch94}, except for the normalisation of the structure
function, so we can compare the results.  Miesch et al.\ (1999)
obtained for several clouds a power law behaviour for the variance of
the centroid velocity differences $\sigma^2(L) \propto L^\gamma$, with
$\gamma\approx 0.85$ (0.33 \dots 1.05) except for the
largest lags, where $\sigma^2$ remained roughly constant. To enable a 
better comparison with the size-linewidth
relation discussed above, we use here the standard deviation $\sigma$
instead of the variance $\sigma^2$.  In Fig.~\ref{fig_variance} we
show the resulting plot for the Polaris Flare data. We find an overall
slope of 0.47, quite close to that found for the size-linewidth
relation. The error bars are somewhat smaller but the two data sets at
higher resolution show some decrease of the slope at the largest lag.  This
must be an artifact due to the finite map size, since it does not
continue at the next larger scale. The structure functions calculated
by \cite{Miesch99} show a much stronger flattening at large lags, going
to constant values for all maps. This is probably due to the
artificial removal of velocity structure at large lags introduced by
their method to subtract large scale trends.

The good agreement between the size-linewidth relation for the
centroid velocities in \S~\ref{sect_larson} and the structure
function discussed here seems inevitable when we consider
that the variance in velocity differences on a certain scale is a kind
of differential measure for the total variance within a certain 
radius as measured with the scanning-beam size-linewidth relation.
We thus expect a similar behaviour.

The kurtosis $K(L)$ is a measure {\changed 
for the correlation of the internal motions.
Values exceeding three at different lags indicate }
the strength of the correlation in velocity space at those
scales. \cite{Miesch99} found that the velocity difference PDFs in the
studied clouds change from kurtosis values between about 10 and 30 at
small lags to nearly Gaussian behaviour at large lags. This is also
typical for incompressible turbulence (\cite{She}). \cite{Lis98} found
strong non-Gaussian distributions at scales associated with filaments
and approximately Gaussians at larger lags.

\begin{figure}
\centering
\epsfig{file=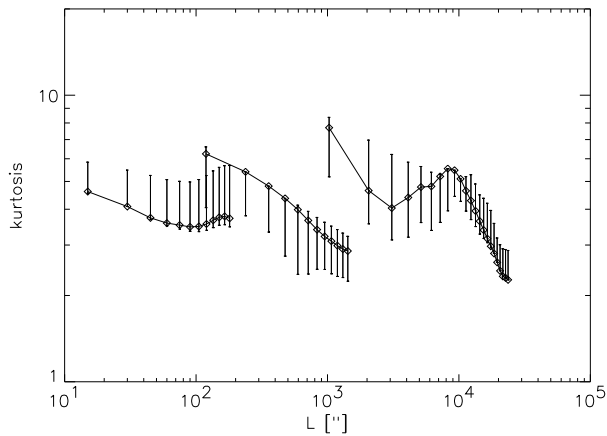,angle=90,width=\columnwidth}
\caption{ The kurtosis of the centroid velocity difference PDF as a
function of the lag between the points considered.}
\label{fig_kurtosis}
\end{figure}

In Fig. \ref{fig_kurtosis} the kurtosis of centroid velocity differences 
in the Polaris Flare data is
plotted. In contrast to the other quantities, the different resolutions do
not line up here to a single line but the kurtosis for each
map drops independently to the Gaussian value at about
the map size. This behaviour can be understood by considering 
which quantities at which scales determine the kurtosis. At the largest 
lag of any map the kurtosis measures mainly the shape of the
centroid probability distribution of the whole map which is more or less 
close to Gaussian for all data sets considered here (\S~\ref{sect_pdfs}). 
Thus we can always
expect a value around 3 when the scale for the kurtosis determination
approaches the map size. Contrary to the discussion provided
by \cite{Miesch99}, this does not mean that there are no correlations 
at larger scales but that they cannot be addressed from points 
within the map.

At all smaller lags the kurtosis is a measure for the correlated motions 
on that scale relative to the overall motions seen in the map, which 
is scanned when computing the kurtosis and variance. In Sect. 
\ref{sect_diffpdfs}
we will see that kurtosis values above three are produced only
if the maps contain some motion on scales larger than the scale
on which the kurtosis is measured. The steps in Fig. \ref{fig_kurtosis} 
are thus unavoidable when switching to another map since we always
measure the correlated motions on a particular scale relative to the 
total motions in the map considered.  
A slightly sub-Gaussian behaviour at the largest scale might be
produced by optical depth effects somewhat flattening the core of the
distribution.

\label{sect_diffpdfs}

\subsection{The $\Delta$-variance}

Stutzki et al. (1998) introduced the $\Delta$-variance to measure the
amount of structure present at different scales in multi-dimensional
data sets.  The $\Delta$-variance at a given scale of an $n$-dimensional
data set is computed by convolving the data with an $n$-dimensional
spherical down-up-down function of that scale, and measuring the
remaining variance.  The $\Delta$-variance analysis computes the
average variance on a certain scale similar to the structure function
giving the variance of the velocity differences between two distinct
points separated by a certain lag.  For the $\Delta$-variance,
however, the variance of the filtered map is computed, instead of the
average variance of all point-to-point differences corresponding to a
certain lag. Thus, the $\Delta$-variance of a smooth map with a linear
gradient vanishes, while the structure function discussed above detects
the gradient. The advantage of the $\Delta$-variance is its better
sensitivity to specific spatial scales. It provides a
good separation of systematic trends, structures on certain scales,
and effects like noise. Furthermore, it allows the direct
computation of the equivalent Fourier
spectral index. A comprehensive discussion is given by Bensch et
al.\ (2001a).  A similar method was introduced recently by Brunt (1999)
to characterise the 3D velocity structure of model
cubes. He used a rectangular filter function composed of adjacent
cubes of different size rather than the spherically symmetric
filter function used to compute the $\Delta$-variance.

Bensch et al.\ (2001a) applied the $\Delta$-variance analysis to the
intensity maps of the Polaris Flare discussed above. We used the same
method in paper~I to analyse the density structure in turbulence
simulations, and compared the results to the observational data.  
Paper I also discussed the $\Delta$-variance for the
3D velocity field of the simulated turbulence and
compared it to the $\Delta$-variance for the 3D
density, but did not provide any direct comparison to the
observations. We found that the $\Delta$-variance of the velocity
behaves similarly to that of the density in showing the
characteristic scale of the driving mechanism used in the turbulence
models. However, the amount of structure observed at smaller scales
differs between density and velocity. The turbulence
creates many thin dense regions, leading to an exponent of the $\Delta$-variance
for the density of about 0.5, whereas it creates hardly any
small-scale structures in the velocity field, so that the
$\Delta$-variance for the velocity drops off much more steeply, with
an exponent of about 2.

Unfortunately this method of measuring the structure of the
3D velocity field in the simulations has no directly
equivalent approach applicable to observations, since they only
provide one-dimensional velocity information projected onto the plane
of the sky. There exists no simple relation between the three dimensional
velocity structure and the behaviour of the projections.
Therefore, we must instead apply the $\Delta$-variance
analysis to observable velocity parameters like the map of centroids
which can be derived from both the simulations and the
observations. They can be used to judge whether a  simulation
reproduces observed properties.  We do lose information by this
procedure, of course.

\begin{figure}
\centering
\epsfig{file=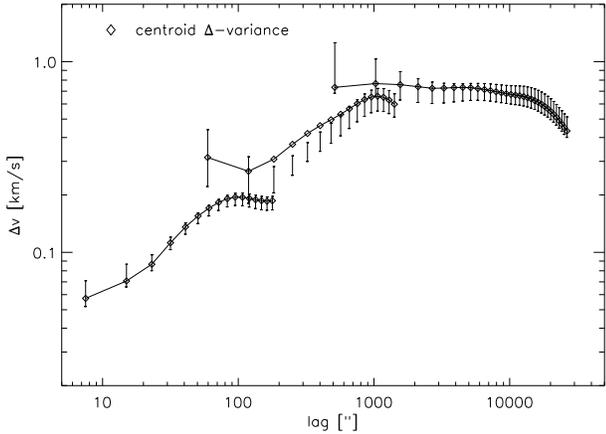,angle=90,width=\columnwidth}
\caption{The square root of the $\Delta$-variance of the velocity
centroid maps of the Polaris Flare observations. {\changed In
contrast to paper~I, we have chosen to take the square root, yielding
the standard deviation of the filtered maps instead of the variance,
in order to give a better comparison to the other linewidth related
quantities.}}
\label{fig_deltavar}
\end{figure}

Fig.~\ref{fig_deltavar} shows the square root of the $\Delta$-variance
for the Polaris Flare velocity centroid maps. The upturn at the
smallest {\changed lags in each map is produced by the observational noise 
adding power at small scales which is only detected in the
$\Delta$-variance spectrum}. \cite{Bensch} showed how the
influence of noise can be subtracted in the $\Delta$-variance. {\newchanged If
we apply this noise correction we find a turn-down
at the smallest lags in each map. This lack of small scale
variations in the different maps is due to the finite
beam size of the observations blurring the structures. 
In the line intensity maps, the influence of the finite beam
can be described by a convolution of the structure with a
Gaussian beam. \cite{Bensch} provided a formalism to correct for both 
the observational noise  and the beam smearing, if one assumes
that the internal scaling of the structure is given by a power law.

In the line centroid maps we cannot apply the same approach
because the beam convolution of the channel maps that is performed
by a telescope does not correspond to the 
convolution of the
line centroid map. Estimates for random velocity fields were
obtained in simulations by \cite{Miesch94}, but they do not
provide a direct way to correct for the beam smearing in the
observed data. To obtain an estimate for the uncertainty introduced
by noise and beam smearing we have tested two extreme approaches.
If we apply the noise correction only and disregard the smallest lags 
in each map, which are obviously influenced by the beam smearing,
we obtain average slopes of 0.04 in the CfA map, 0.46 in the KOSMA
map, and 0.68 in the IRAM map. In this case the curves do not line up
perfectly. The $\Delta$-variance for one lag in a larger-scale
map always falls somewhat below the $\Delta$-variance for the same lag
in the smaller-scale map, indicating that the slopes are probably 
overestimated. If, on the other hand, we assume that the beam provided
a simple convolution 
of the centroid map, so that we can apply Eq. (15) from \cite{Bensch},
we obtain slopes of -0.05, 0.36, and 0.32, respectively and
the opposite misalignment. The true slopes should fall
in between these two extreme cases. } 



Over most scales covered by the CfA map, the slope flattens to zero,
in contrast to the behaviour in the intensity maps.  This virtual lack
of large-scale variations does not reflect the real structure but
appears {\changed to be} due to the missing weighting in the
$\Delta$-variance analysis.  Here, all points are counted equally,
even those with intensities well below the noise limit. Thus, the
$\Delta$-variance analysis necessarily fails in cases where the maps
are only sparsely filled by emission.  Less than one third of the CfA
map shows emission above the noise limit so that we cannot expect any
significant results at scales above half a degree. 

{\newchanged For the maps at smaller scales without large ``empty'' 
regions, we can expect significant $\Delta$-variance spectra  but
we must admit that without a good theory for the influence of the 
beam convolution on line centroids we can hardly separate the effects
of the beam smearing from the velocity scaling in maps which
are not large compared to the beam size. Thus large error bars remain 
for the measured velocity scaling in the Polaris Flare preventing 
any definite conclusion on the scale dependence of the slope in the 
velocity structure. 

In contrast the size-linewidth relation and the structure function
do not show clear signatures of noise and beam smearing in Figs.
\ref{polaris_larson} and \ref{fig_variance}. To test their
 sensitivity we have applied them to data smoothed
 with a Gaussian filter tuned to ensure that the corresponding
$\Delta$-variance shows the same behaviour as the
$\Delta$-variance where the noise was removed following \cite{Bensch}.
Only small changes relative to the original data are seen. The
size-linewidth relation and the structure function of the velocity
centroids exhibit a weak steepening of the slope by about 0.04, while
the velocity PDFs and the kurtosis show no clear differences.
The comparison of the results from the original maps and the
smoothed data shows that these other methods are not particularly
sensitive to uncorrelated noise or beam smearing at small scales.}

\subsection{Comparison of the methods}

We can classify all of the methods we have described in terms of the
velocity information setup in a two-dimensional map, the filtering
function used, and the weighting of the data in the map.

Most analyses were restricted to the velocity centroids, which
are effectively the first moment of the local velocity profile. The
size-linewidth relation adds the local variance, i.e. the
second velocity moment, and the study of the PDFs also uses
the kurtosis.
However, with sufficiently high signal-to-noise,
higher moments may provide valuable additional information.

The variance is strongly dominated by the depth of the observed cloud,
so that it contains information lost when considering the velocity 
centroids only. We have seen that, for maps where the line profiles 
sample the cloud deeply in comparison to
the map size, the integrated line profile is a better measure for the
true velocity distribution function than the PDF of the velocity
centroids.

We have applied three different kinds of filters: the scanning-beam
size-linewidth relation effectively convolves the map with a positive
Gaussian filter; the $\Delta$-variance analysis uses a spherically
symmetric up-down filter; and the structure function uses a filter
consisting of a positive and a negative spike separated by a certain
distance. In the latter case, spherical symmetry is provided by the
superposition of the resulting variance values for different
directions of the filter axis.  The structure function is sensitive to
large-scale gradients and can detect certain geometric structures, but
because of the strong localisation of the filter in the spatial
domain, it is unfortunately sensitive to a broad spectrum of spatial
frequencies in the Fourier domain. In the statistical analysis of
velocity fluctuations, it is therefore at a disadvantage in the
detection of characteristic scales and frequencies compared to the
$\Delta$-variance analysis. Similar conclusions were obtained by
\cite{Houlahan}. {\newchanged This may explain why we detect the 
scale of noise only by means of the $\Delta$-variance, whereas its
influence is hidden in the full spectra of the structure function
and the size-linewidth relation.}

The $\Delta$-variance analysis, on the other hand, does not yet take
into account different weights for the information in different
regions of an observed map, so that it fails for maps with large
regions dominated by noise.  The weighting of the velocity centroid
information by the intensity, as is done automatically in the
size-linewidth relation, reduces the uncertainty due to
observational noise when computing the centroid probability
distribution or structure function. 

\label{sect_comparemethods}

\subsection{Other approaches}

With sufficiently high signal-to-noise ratio, it is possible
to extend the methods discussed here.  Overviews of the
different existing methods have been provided recently by V\'azquez-Semadeni
(2000) and Ossenkopf et al.\ (2000).  First,
one can apply the basically two-dimensional methods to higher
moments of the line profiles, providing new information especially
on the intermittency in velocity space.
Alternatively, the velocity channel maps  can be analysed
as demonstrated with the $\Delta$-variance analysis by
Ossenkopf et al.\ (1998). Another method is to compare full spectra, using
the spectral correlation function  (Rosolowsky
et al.\ 1999) and extending this method to consider all spatial variations. 

Tauber (1996) discussed the smoothness of line profiles as a measure for 
the size and number of coherent units contributing to the profiles.
Applying a rough approximation to this analysis, Falgarone et al.\ 
(1998) conclude that the size of cells in the Polaris Flare observations 
must be as low as 200~AU.

When looking for characteristic global features in the
density-velocity structure, the principal component analysis introduced
by Heyer \& Schloerb (1997) is probably the most significant tool.  It
identifies the main components in the position-velocity space in terms
of eigenvectors and eigenimages.  Although the principal component
analysis represents a reliable method to find the dominant 
structures even in complicated images, the significance of the
higher-order moments still has to be determined. 

\section{Turbulence models}

\subsection{Simulations}

We use simulations of uniform decaying or driven turbulence with and
without magnetic fields described by Mac Low et al.\ (1998) in the
decaying case and by Mac Low (1999) in the driven case.  These
simulations were performed with the astrophysical MHD code
ZEUS-3D\footnote{Available from the Laboratory for
Computational Astrophysics of the National Center for Supercomputing
Applications, http://zeus.ncsa.uiuc.edu/lca\_home\_page.html} 
(\cite{c94}).  This is a 3D version of the code
described by Stone \& Norman (1992a,b) using second-order advection
(\cite{v77}), that evolves magnetic fields using constrained
transport (\cite{eh88}), modified by upwinding along shear
Alfv\'en characteristics (\cite{hs95}).  The code uses a von
Neumann artificial viscosity to spread shocks out to thicknesses of
three or four zones in order to prevent numerical instability, but
contains no other explicit dissipation or resistivity.  Structures
with sizes close to the grid resolution are subject to the usual
numerical dissipation, however. 
In Paper I we discussed the effects of limited
numerical resolution, which leads to numerical viscosity, and noted
that resolution studies could be used to determine which properties
were well resolved.

The simulations used here were performed on a 3D,
uniform, Cartesian grid with side $L = 2$ and
periodic boundary conditions in every direction, using an isothermal
equation of state. To deal with velocities comparable to those in the
observations we have assumed here a cloud temperature of 10~K
corresponding to a translation of the dimensionless sound speed 
in the simulations to a physical sound speed of 0.2~km/s in the
data analysis. The initial density and, in relevant cases, magnetic
field are both initialised uniformly on the grid, with the initial
density $\rho_0 = 1$ and the initial field parallel to the $z$-axis.
The turbulent flow is initialised with velocity perturbations drawn
from a Gaussian random field determined by its power distribution in
Fourier space, as described by Mac Low et al.\ (1998).  For decaying
models we use a flat spectrum with power in the range $k\sub{d} = 1-8$, where
the dimensionless wavenumber $k\sub{d} = L/\lambda_d$ counts the number of
driving wavelengths $\lambda_d$ in the box.  For our driven models, we
use a spectrum consisting of a narrow band of wave numbers around some
value $k\sub{d}$, and driven with a fixed pattern at constant kinetic energy
input rate, as described by Mac Low (1999).

We have tested the influence of numerical viscosity by running the
simulations with the same physical parameters on grids of 64$^3$,
128$^3$ or 256$^3$ zones. Higher resolution grids have numerical
viscosity acting at smaller scales, so changing the resolution shows
the effects of numerical viscosity on our results. 
The influence of numerical resolution on the simulation results 
is discussed below separately for each for the statistical measures.

\subsection{Simulated Observations}

To compare our simulations with observations we must synthesise
observational maps from the simulated density and velocity fields.  We
assume that the cubes are optically thin for this first study so that
direct integration along lines of sight through the cube neglecting
optical depth effects yields line profiles.  This appears to be a
reasonable assumption for comparison with the low column density
clouds in the Polaris flare observed in $^{13}$CO, but is a
worse assumption for higher column density clouds or more optically
thick species.  

We also neglect the periodic nature of the simulations, effectively
observing the simulation cubes as isolated structures in a vacuum.
This second assumption must be taken into account in analyses affected
by the path-length through the cloud, such as comparisons of the
velocity PDF measured from the average line profile vs.\ the centroid
velocity distribution.

\subsection{Statistical fluctuations}

\begin{figure}
\centering
\epsfig{file=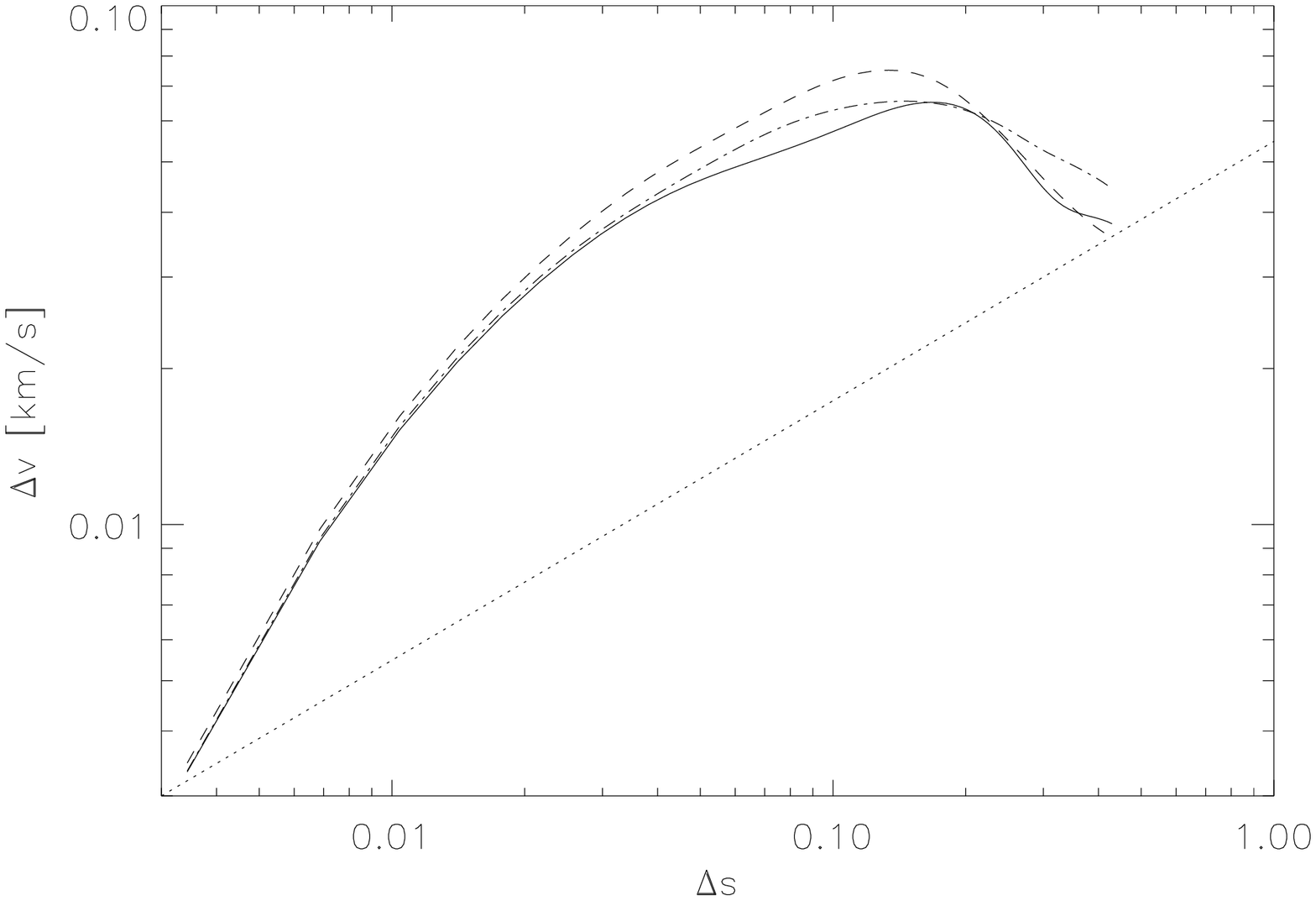,angle=90,width=\columnwidth}
\epsfig{file=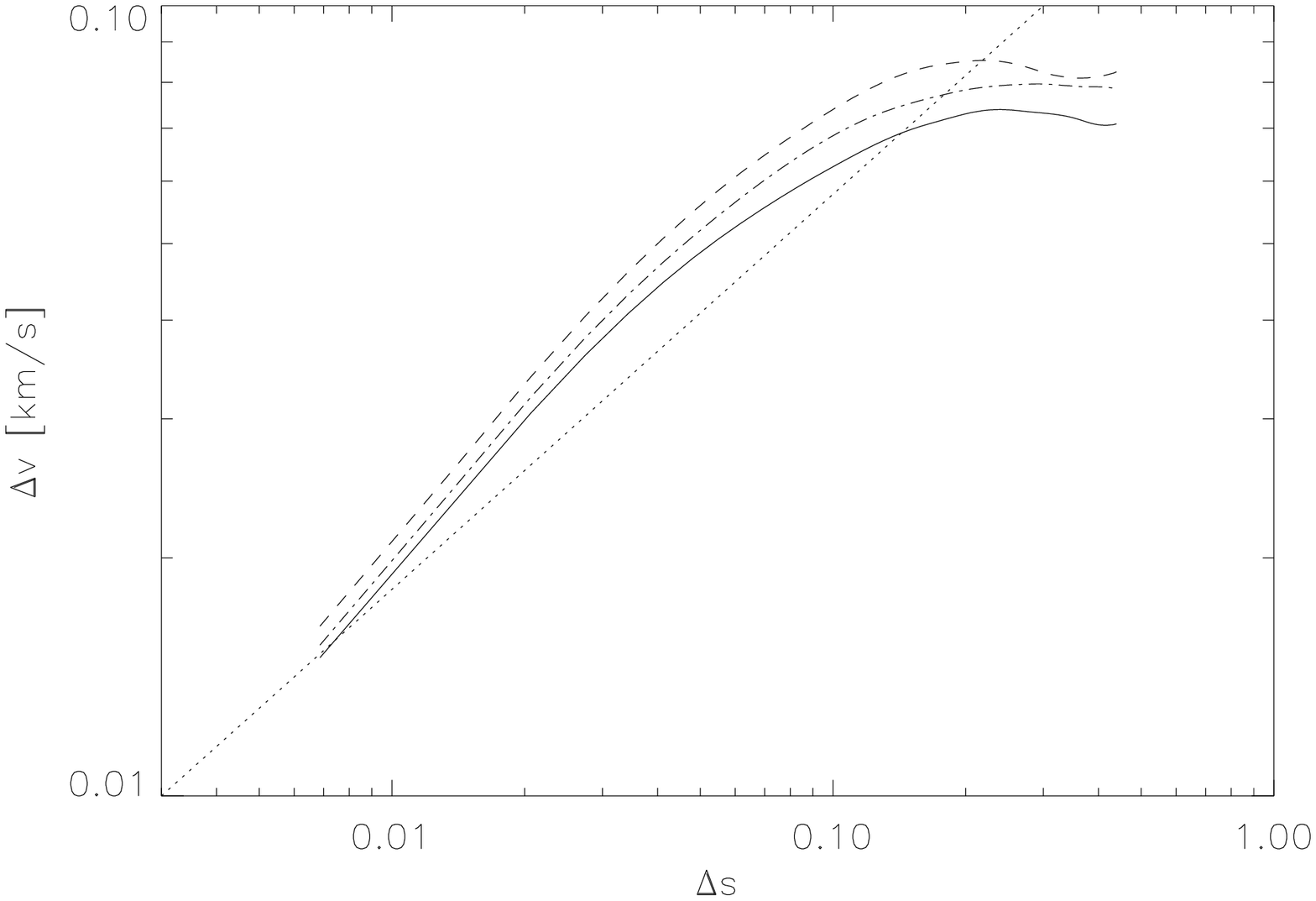,angle=90,width=\columnwidth}
\caption{$\Delta$-variances (upper panel) and structure
functions (lower panel) in the centroid velocity 
maps obtained for a decaying hydrodynamic model (model D
from Mac Low et al.\ 1998) in the three
possible directions of projection. The dotted line indicates
a slope of 0.5 for comparison.  The variation between the 
different directions is similar for the other models. In the
$\Delta$-variance plots we always give the standard deviation (the
square root of the $\Delta$-variance) for a
better comparison to the other velocity variations.}
\label{fig_directions}
\end{figure}

To get a feeling for the significance of the structural properties
indicated by the different measures relative to the statistical
variations in the turbulence, {\changed which may build up from isotropic
initial perturbations during the turbulence evolution,} we can compare 
different directions within the same model cube.

Fig. \ref{fig_directions} shows the $\Delta$-variances and structure
functions for the three centroid velocity maps 
{\changed along the three axes in a decaying hydrodynamic model.} 
The $\Delta$-variances show {\changed variations of 40\% in value at the
largest scales,}
where the exact modal structure is still different in the 
three directions. 
The structure function shows smaller variations corresponding
to its lower sensitivity to changes at particular spatial {\changed scales.}
The variation of the slope in the different directions falls below
0.1.  The equivalent plot for the size-linewidth relation is similar
to that for the structure function. The variations in the kurtosis,
on the other hand, are more like those seen in the
$\Delta$-variance. Here, we sometimes find small dips and rises
distorting the monotonic decay to values slightly below 3 at the
largest scales.
In the total and the centroid velocity PDFs we find {\changed substantial} 
variations in the core and the central position of the distributions 
corresponding to the different largest velocity modes but no
changes in the wing behaviour.


\section{Statistical description of simulations and comparison to observations}

\label{sect_sims}
\subsection{Size-linewidth relation}

\begin{figure}
\centering
\epsfig{file=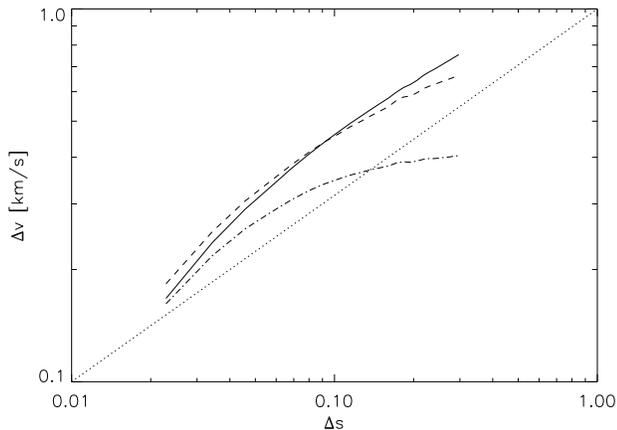,angle=90,width=\columnwidth}
\caption{Size-linewidth relation for
the velocity centroids of a hydrodynamic simulation driven
at wavenumbers of $k\sub{d}=2$ (solid), $k\sub{d}=4$ (dashed), and $k\sub{d}=8$
(dot-dashed).  These are models HE2, HE4, and HE8 from
\cite{ml99}. The dotted line shows a power-law with slope 0.5.}
\label{fig_newlars}
\end{figure}

We begin by considering the results of applying the size-linewidth
analysis described in \S~\ref{sect_larson} to the models.  In
Fig.~\ref{fig_newlars} we show the size-linewidth relation for the
velocity centroids in three models of hydrodynamic driven turbulence
that differ only in the scale that they are driven. The driving wavelengths
are 1/2, 1/4, and 1/8, respectively.

We find power-law behaviour through most of the regime only for the
model driven at the largest available scales. Models driven with
smaller characteristic scales show a flattening of the relation at
lags above the driving scale. A slight flattening at the largest 
lags is also visible in the observational data. This appears to be 
an indication of a turbulence driving scale close to the size of
the molecular cloud.

\begin{figure}
\centering
\epsfig{file=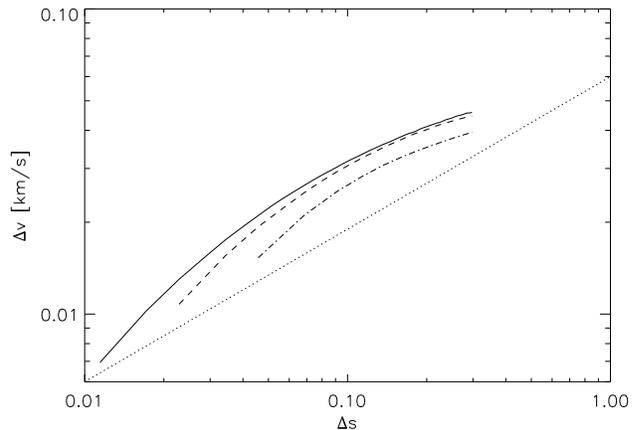,angle=90,width=\columnwidth}
\caption{Size-linewidth relation for
the velocity centroids in a hydrodynamic simulation of
decaying turbulence at time $t = 0.75 t_{\rm cross}$ 
($t_{\rm cross}$ is the initial crossing time of
the region), computed at resolutions of $64^3$ (dot-dashed), 
$128^3$ (dashed), and $256^3$ (solid). 
These are models B, C, and D from \cite{ml98}.}
\label{fig_newlarsres}
\end{figure}
A drop off in velocity dispersion is seen at small lags in all of the
models.  This can be explained straightforwardly as an effect of
numerical viscosity.  In Fig.~\ref{fig_newlarsres} we show a
comparison of three models of decaying turbulence that are
statistically identical, but were computed at resolutions of $64^3$,
$128^3$, and $256^3$ zones. 
Increasing resolution results in decreasing numerical viscosity, so
Fig.~\ref{fig_newlarsres} demonstrates explicitly the effect of
changing the numerical viscosity.  As can be seen, the slope does not
change at large lags, and the higher resolution models agree within a
few percent on the magnitude of the velocity dispersion. At small
lags, on the other hand, the velocity dispersion falls off.  This 
occurs in all models at roughly the same number of grid zones,  and
hence at larger physical scale in the lower resolution
models. This part of the spectrum thus reflects the effect of
numerical diffusion eliminating some of the small-scale structure.
The same drop-off occurs in each model (see Fig.~\ref{fig_newlars}), 
so the behaviour at
the smallest scales should be viewed with caution. We note, however,
that a physical diffusion such as ambipolar diffusion (Zweibel \&
Josafatsson 1983, Klessen et al.\ 2000) is expected to produce similar
behaviour at the diffusion scale.

Magnetic fields do not appear to modify the size-linewidth relation, although
they can make order unity differences in the magnitude of the velocity
dispersion and produce significant anisotropy, as described
below in \S~\ref{sim_delta}. 

We can conclude that the observed power-law
behaviour of the size-linewidth relation is reasonably explained by
either hydrodynamic or magnetised turbulence driven at scales
comparable to the largest observed scales. 

\subsection{Velocity probability distribution function}
\label{sect_simpdfs}


\subsubsection{Centroid velocity PDFs}
\label{subsub-cenvelpdf}
\begin{figure}
\centering
\epsfig{file=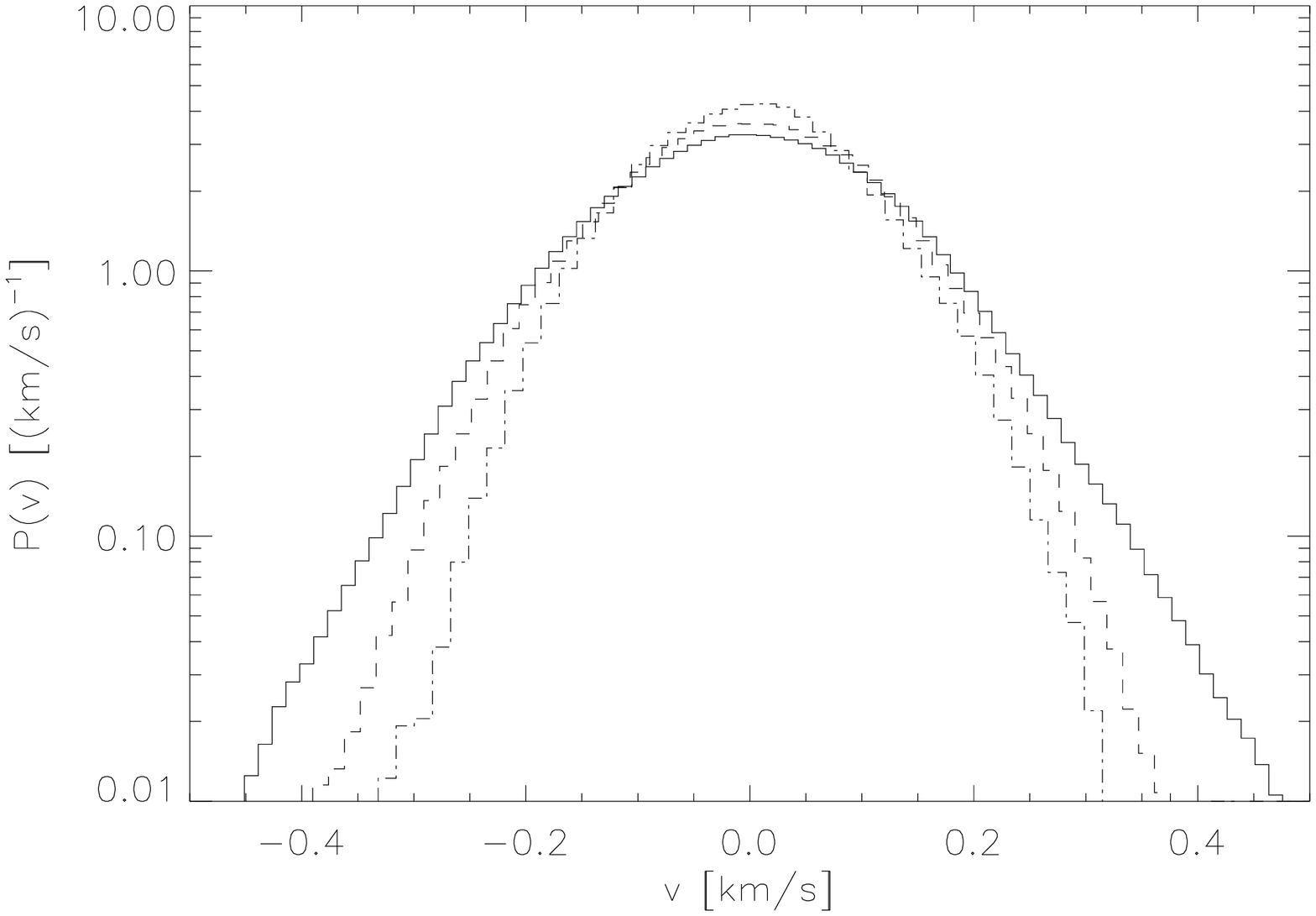,angle=90,width=\columnwidth}
\epsfig{file=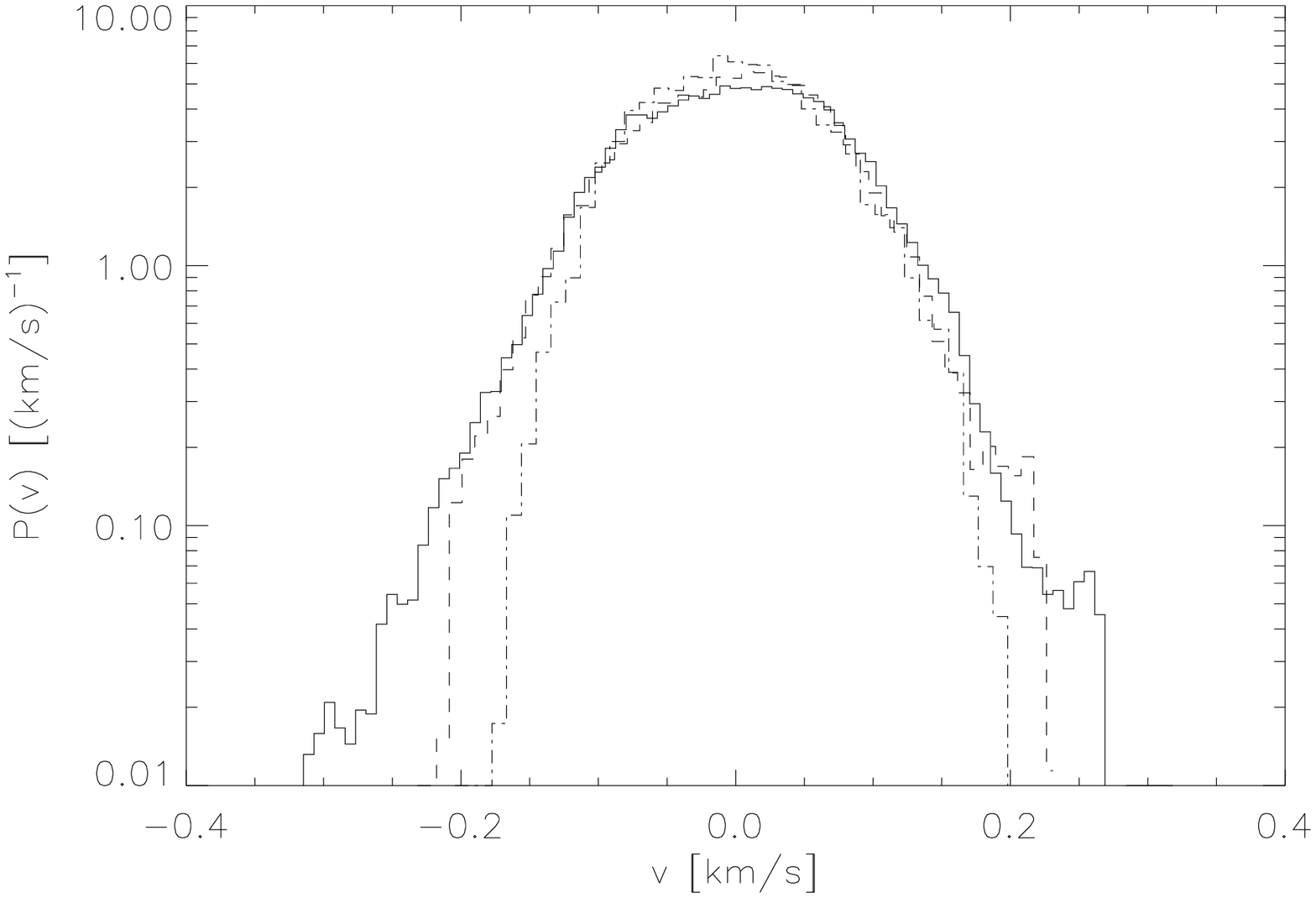,angle=90,width=\columnwidth}
\caption{(a) 3D and (b) centroid velocity PDFs in a model of strongly
magnetised decaying turbulence with initial Mach number $M = 5$ and Alfv\'en
number $A = 1$
after $1.5 t_{\rm cross}$ at a resolution of 256$^3$ (solid), 128$^3$
(dashed), and 64$^3$ (dashed-dotted). (Models Q, P, N from Mac Low
et al. 1998)}
\label{fig_pdfresolv}
\end{figure}

{\changed We begin our study of velocity PDFs in the models} by
examining the effect of numerical resolution on the centroid velocity
PDFs.  In Fig.~\ref{fig_pdfresolv} the 3D PDFs are compared to the
centroid PDFs for a model of decaying MHD turbulence at different
resolutions. All PDFs are well represented by Gaussians. The width of
the PDF drops at smaller resolution.  This can be attributed to the
stronger influence of the numerical viscosity in the lower resolution
cubes damping the turbulence faster. This agrees with the measurements
of resolution effects on kinetic energy described in \cite{ml98}, in
which the magnitude of the kinetic energy increased with resolution,
although the decay rate was constant.  An additional effect is seen in
the centroid PDFs. At low resolutions the sampling of the wings of the
Gaussian is insufficient, so that the distribution appears too
narrow. The kurtosis values of these PDFs are 2.6, 2.8, and 2.9 with
growing resolution. Hence, sub-Gaussian kurtosis values can at least
partially be explained by small map sizes.

\begin{table*}
\caption{Parameters of the centroid velocity PDFs and the total
velocity PDFs for the model cubes
\label{tab-modelpdfs}}
\begin{tabular}[htp]{lllllllllllll}
model$^{\mathrm a}$ & $L^{\mathrm b}$ & $k_d^{\mathrm c}$  & $N^{\mathrm d}$ 
             & $M^{\mathrm e}$ & $\frac{v_A}{c_s}^{\mathrm f}$ &
             $t^{\mathrm g}$ & 
             $\frac{\sigma^2}{c_s^2}$(cen)$^{\mathrm h}$  &  
             $\frac{\sigma^2}{c_s^2}$(cube)$^{\mathrm h}$ & 
             $\frac{\sigma^2(\mbox{cube})}{\sigma^2(\mbox{cen})}$
             & $K$(cen)$^{\mathrm j}$ & $K$(cube)$^{\mathrm j}$  \vspace{0.5mm}\\
\hline	
HA8   & 0.1 & 7--8 & 128 & 1.9 & 0 & 1.0 & 3.6   & 10  & 2.8 & 3.0 &  2.9 \\
HC2   & 1   & 1--2 & 128 & 7.4 & 0 & 1.0 & 50    & 55  & 1.1 & 2.4  &  2.5 \\
HC4   & 1   & 3--4 & 128 & 5.3 & 0 & 1.0 & 17    & 31  & 1.8 & 3.2  &  2.7 \\
HC8   & 1   & 7--8 & 128 & 4.1 & 0 & 1.0 & 9.4   & 22  & 2.3 & 3.6  &  3.0 \\
HE2   & 10  & 1--2 & 128 & 15  & 0 & 0.98& 49    & 76  & 1.6 & 3.0  &  2.7 \\
HE4   & 10  & 3--4 & 128 & 12  & 0 & 0.88& 38    & 67  & 1.8 & 2.9  &  3.0 \\
HE8   & 10  & 7--8 & 128 & 8.7 & 0 & 1.0 & 21    & 47  & 2.2 & 3.9  &  3.0 \\
MA4X: $v_{\bot}$
      & 0.1 & 3--4 & 128 & 2.7 & 10& 0.3 & 7.0   & 14  & 2.0 & 2.6  &  2.8 \\
\hspace{3em} $v_{\|}$    
      & 0.1 & 3--4 & 128 & 2.7 & 10& 0.3 & 10    & 16  & 1.6 & 3.4  &  3.3 \\
MC4X: $v_{\bot}$
      & 1   & 3--4 & 128 & 5.3 & 10& 0.1 & 15    & 26  & 1.7 & 3.8  &  3.6 \\
\hspace{3em} $v_{\|}$
      & 1   & 3--4 & 128 & 5.3 & 10& 0.1 & 18    & 29  & 1.6 & 3.3  &  3.3 \\
MC45: $v_{\bot}$
      & 1   & 3--4 & 128 & 4.8 & 5 & 0.2 & 13    & 23  & 1.8 & 4.1  &  3.4 \\
\hspace{3em} $v_{\|}$
      & 1   & 3--4 & 128 & 4.8 & 5 & 0.2 & 16    & 29  & 1.8 & 3.1  &  3.0 \\
MC41: $v_{\bot}$
      & 1   & 3--4 & 128 & 4.7 & 1 & 0.5 & 15    & 28  & 1.9 & 3.0  &  2.8 \\
\hspace{3em} $v_{\|}$
      & 1   & 3--4 & 128 & 4.7 & 1 & 0.5 & 13    & 24  & 1.8 & 3.1  &  3.0 \\
MC85: $v_{\bot}$
      & 1   & 7--8 & 128 & 3.4 & 5 &0.075& 6.3   & 15  & 2.4 & 4.1  &  3.8 \\
\hspace{3em} $v_{\|}$
      & 1   & 7--8 & 128 & 3.4 & 5 &0.075& 9.0   & 21  & 2.3 & 3.2  &  3.5 \\
MC81: $v_{\bot}$
      & 1   & 7--8 & 128 & 3.5 & 1 & 0.23& 6.6   & 19  & 2.9 & 3.3  &  3.1 \\ 
\hspace{3em} $v_{\|}$
      & 1   & 7--8 & 128 & 3.5 & 1 & 0.23& 8.5   & 20  & 2.4 & 3.3  &  3.2 \\ 
ME21$^{\mathrm k}$: $v_{\bot}$
      & 10  & 1--2 & 128 & 14  & 1 & 0.5 & 47    & 61  & 1.3 & 2.8  &  2.9 \\
\hspace{3em} $v_{\|}$  
      & 10  & 1--2 & 128 & 14  & 1 & 0.5 & 83    & 97  & 1.2 & 3.3  &  3.2 \\
B     & 0   & 1--8 & 64  & 5   & 0 & 0.1 & 2.7   & 7.0 & 2.6 & 2.6 &  2.9 \\
      & 0   & 1--8 & 64  & 5   & 0 & 0.3 & 1.9   & 4.0 & 2.1 & 2.8 &  2.9 \\
C     & 0   & 1--8 & 128 & 5   & 0 & 0.1 & 2.8   & 7.7 & 2.8 & 3.0 &  3.1 \\
      & 0   & 1--8 & 128 & 5   & 0 & 0.3 & 1.7   & 4.4 & 2.6 & 3.2 &  2.9 \\
      & 0   & 1--8 & 128 & 5   & 0 & 0.5 & 1.3   & 3.5 & 2.7 & 2.7 &  2.7 \\ 
D     & 0   & 1--8 & 256 & 5   & 0 & 0.1 & 2.9   & 8.4 & 2.9 & 3.0 &  2.9 \\
      & 0   & 1--8 & 256 & 5   & 0 & 0.3 & 1.9   & 5.0 & 2.6 & 2.9 &  2.9 \\
      & 0   & 1--8 & 256 & 5   & 0 & 0.5 & 1.4   & 3.8 & 2.7 & 2.4 &  2.9 \\ 
U$^{\mathrm k}$
      & 0   & 1--8 & 256 & 50  & 0 & 0.1 & 9.1   & 15  & 1.6 & 2.3 &  2.9 \\
      & 0   & 1--8 & 256 & 50  & 0 & 0.3 & 6.1   & 8.9 & 1.5 & 2.1 &  2.6 \\
      & 0   & 1--8 & 256 & 50  & 0 & 0.5 & 5.0   & 7.0 & 1.4 & 2.1 &  2.5 \\
Q: $v_{\bot}$ 
      & 0   & 1--8 & 256 & 5   & 5 & 0.1 & 3.6   & 11  & 3.1 & 3.2 &  3.3 \\
      & 0   & 1--8 & 256 & 5   & 5 & 0.3 & 2.4   & 6.4 & 2.7 & 2.9 &  3.3 \\
      & 0   & 1--8 & 256 & 5   & 5 & 0.5 & 2.1   & 4.9 & 2.3 & 3.0 &  3.2 \\
Q: $v_{\|}$
      & 0   & 1--8 & 256 & 5   & 5 & 0.1 & 4.7   & 9.5 & 2.0 & 2.9 &  3.0 \\
      & 0   & 1--8 & 256 & 5   & 5 & 0.3 & 3.9   & 6.3 & 1.6 & 3.0 &  3.4 \\
      & 0   & 1--8 & 256 & 5   & 5 & 0.5 & 3.4   & 4.9 & 1.4 & 3.1 &  3.3 \\  
\hline
\label{tab_modelpdfs}
\end{tabular}
\\ \footnotesize
${}^{\mathrm a}$ Driven models (H \& M series) from \cite{ml99};
decaying models (single letters) from \cite{ml98}. \\ 
${}^{\mathrm b}$ Mechanical driving luminosity in arbitrary units (see
\S~2.3 of \cite{ml99} for all unit conversions). \\
${}^{\mathrm c}$ Driving wavenumber. \\
${}^{\mathrm d}$ Number of zones in each dimension. \\
${}^{\mathrm e}$ rms Mach number: initial value for decaying models,
equilibrium value for driven models. \\
${}^{\mathrm f}$ Ratio of Alfv\'en velocity to sound speed. \\
${}^{\mathrm g}$ Time at which values are measured, in sound-crossing times.\\
${}^{\mathrm h}$ Variance of distribution for line centroid velocities
and full cube. \\
${}^{\mathrm j}$ Kurtosis of distribution for line centroid velocities
and full cube. \\
${}^{\mathrm k}$ Unpublished model. \\
\normalsize
\end{table*}

In Table~\ref{tab-modelpdfs} we give the PDF moments for most of the
models discussed, covering a wide range of different physical properties.
The first several columns describe the model input parameters, 
and the remaining columns contain the parameters of the PDFs obtained.

\begin{figure}
\centering
\epsfig{file=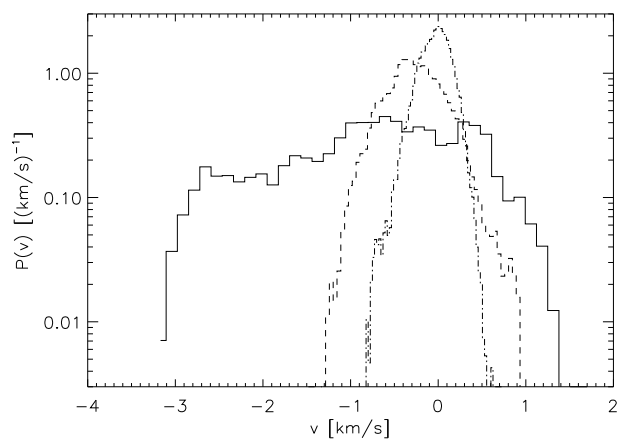,angle=90,width=\columnwidth}
\caption{Centroid velocity PDFs from  three models driven with the
same strength at wavenumbers of $k\sub{d}=2$ (solid), $k\sub{d}=4$ (dashed),
and $k\sub{d}=8$ (dash-dotted) (models HC2, HC4, and HC8 from Mac Low
1999)
\label{fig-newcvpdf}
}
\end{figure}
We first consider the effect of varying the driving wavenumber, holding the
energy input constant. Fig.~\ref{fig-newcvpdf} and the corresponding
values in Table~\ref{tab-modelpdfs}
show that driving at smaller wavenumbers (longer wavelengths) produces
broader PDFs, because such models have lower dissipation rates (Mac Low 1999), 
and thus higher rms velocities.  More interestingly, we find that driving at
the largest scale in the model $k\sub{d}=2$ produces a centroid velocity
PDF with apparently non-Gaussian shape, as reflected in the kurtosis
value of 2.4 given in Table~\ref{tab-modelpdfs}.  A similar
result was found by Klessen (2000), who argued that it is most likely
due to cosmic variance.  That is, an insufficient number of modes are
sampled at these long driving wavelengths to fully describe a
Gaussian field, so the PDF appears to have a distorted shape. 
Depending on the random numbers used to initialise the largest modes,
both Gaussian and non-Gaussian kurtosis values are then possible.

\begin{figure}
\centering
\epsfig{file=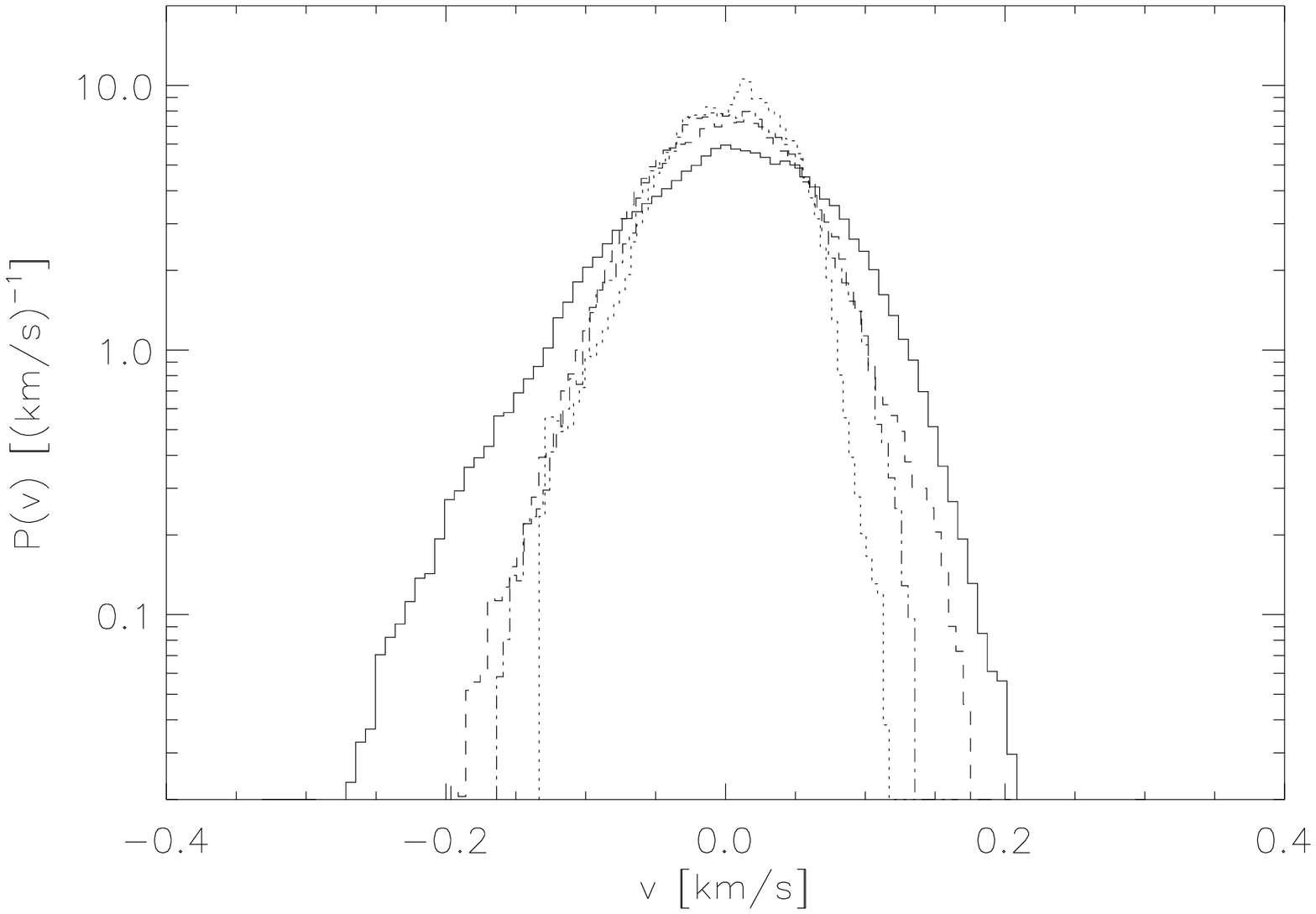,angle=90,width=\columnwidth}
\epsfig{file=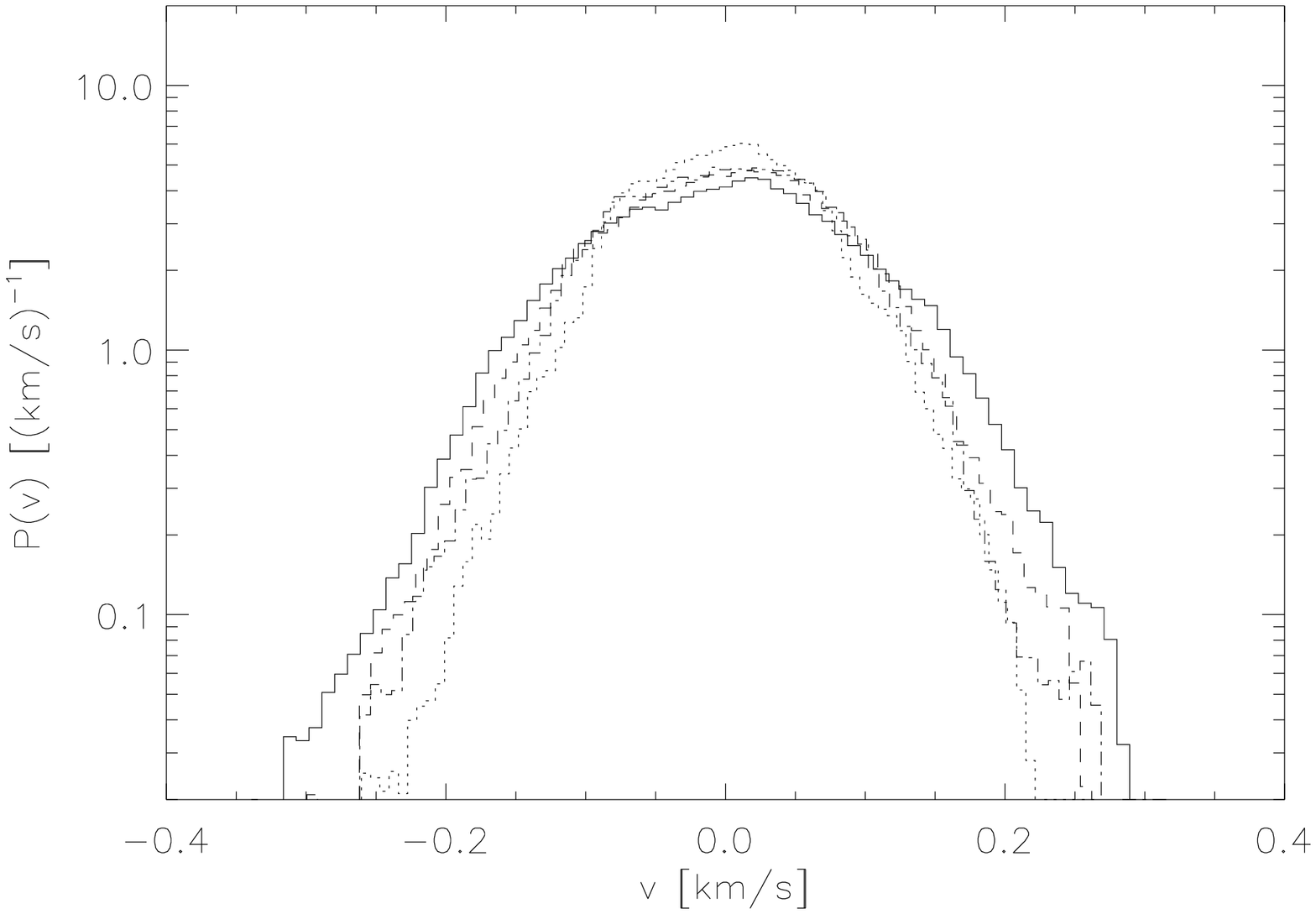,angle=90,width=\columnwidth}
\caption{Centroid velocity PDFs from a model of decaying, magnetised
turbulence (model Q from Mac Low et al.\ 1998) at times in units of the initial
crossing time $t_{\rm cross}$ of 0.5 (solid), 0.75 (dashed),1.5
(dash-dot), and 2.5 (dotted) observed (a)
perpendicular to the field and (b) parallel to the field.
\label{fig-newmagdecaypdf}
}
\end{figure}

To demonstrate the effect of magnetic fields,
Fig.~\ref{fig-newmagdecaypdf} shows the centroid velocity PDFs from a
model of decaying magnetised turbulence. It clearly indicates an 
anisotropic decay, with velocity components perpendicular to
the magnetic field decaying substantially more quickly than 
velocities parallel to the field.  In both cases, though, the
PDFs remain Gaussian even at late times.  These conclusions are
quantitatively supported by Table~\ref{tab-modelpdfs}.

{\changed We also examined magnetised driven turbulence.
The fields do slightly shift} the peak, but, as
shown in Table~\ref{tab-modelpdfs}, they still produce only 
{\changed marginally}
non-Gaussian centroid velocity PDFs in cases where the hydrodynamic
model is Gaussian, contrary to the speculation of Klessen (2000) that
magnetic fields might be an important alternative cause of
non-Gaussian PDFs.  The PDFs observed parallel to the field
are roughly 20\% wider than the perpendicular observations {\changed
as seen in  Table~\ref{tab-modelpdfs}.}

Klessen (2000) showed that driving from large scales can produce
non-Gaussian PDFs, but worried that every additional piece of physics
also appeared likely to produce non-Gaussian PDFs, allowing no
conclusions to be drawn from their occurrence.  We have, however,
demonstrated that neither magnetic fields nor the vorticity introduced
by shock interactions in driven or decaying turbulence produce
strongly non-Gaussian PDFs.  

{\changed Therefore, the non-Gaussian PDFs observed in the Polaris
Flare (\S~\ref{subsect-vpdf}) must have a different explanation.}
Another candidate for producing non-Gaussian PDFs is self-gravity, but
the lack of star-forming activity in the Polaris Flare suggests that
self-gravitation does not play a dominant role there.  This suggests
that the non-Gaussian PDFs observed there are indeed due to the cosmic
variance introduced by driving from the largest scales of the region.
The driving scale may actually be even larger than identified here
because we have not used any information from the atomic gas at larger
scales.

\subsubsection{Average line profiles}

Using the turbulence simulations we can now revisit the question
from \S~\ref{sect_pdfs} of how best to measure the actual 3D 
velocity PDF from observations.
\begin{figure}
\centering
\epsfig{file=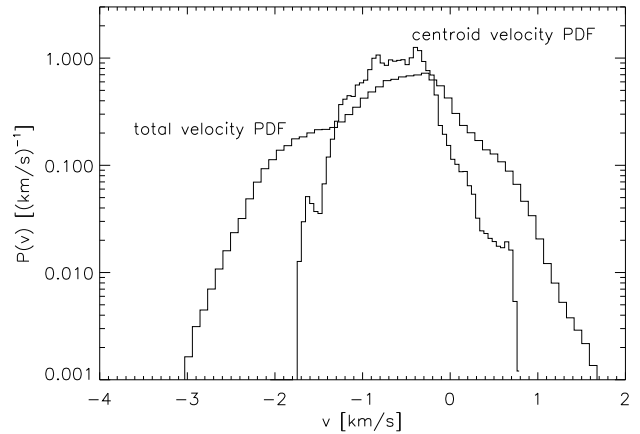,angle=90,width=\columnwidth}
\caption{Comparison of the total velocity PDF and the
centroid velocity PDF for a hydrodynamic model driven with
one tenth of the energy of the example above (model HC2 from
Mac Low 1999).}
\label{fig_centpdfs}
\end{figure}
In Fig. \ref{fig_centpdfs} we compare the PDF of the whole velocity
distribution that would be measured as the average line profile 
in an optically thin LTE medium 
to the PDF of the centroid velocities for a hydrodynamic model driven
at $k\sub{d}=2$.  This plot may be compared with
Fig.~\ref{fig_irampdfs}, showing the centroid velocity distribution and
the average line profile measured for the IRAM map in the Polaris
Flare.

In both cases we find similar distorted Gaussian distributions, with
the centroid velocity distribution narrower than the full velocity
distribution.  Table~\ref{tab-modelpdfs} contains the ratio between
the widths of the two distributions for all models.  The
model ratios show substantial variation around a typical
value of about two. The strongest systematic variation appears to be
with initial or driven rms Mach number $M_0$, with higher $M_0$ giving 
lower ratios, down to as low as 1.1 for the most strongly driven
model HE2. 
Observing parallel to the field lines leads to somewhat lower ratios
than perpendicular, presumably due to the higher velocity variance
seen along parallel lines of sight.
However, the ratio also decreases during the decay of turbulence for
unclear reasons.

The ratio between depth and width is fixed to unity in the model.
From the combination of the three Polaris Flare maps we estimate in
\S~\ref{sect_pdfs} a typical width ratio between 1.5 and 1.6 for a
depth comparable to the lateral extension.  In
Fig. \ref{fig_centpdfs}, showing a very high $M_0$ simulation, the
ratio is 1.5, but for the majority of simulations we find ratios of
over 1.8, suggesting that the {\changed Polaris Flare} observations
are of a region containing hypersonic turbulence with $M_0$ of order
10 or higher.

For most model PDFs we have tested three possible fits to the distribution:
Gaussians fitting either the whole distribution or only the wings,
and an exponential fit. Although exponential wings cannot be ruled 
out completely, the Gaussian fits are clearly better. We have found 
this to be true for all models of driven and decaying turbulence
discussed here, as can be seen by visual inspection of 
Figs. \ref{fig_pdfresolv} to \ref{fig_centpdfs}. This result agrees 
with the Gaussian velocity PDFs found by \cite{Chappell} for models 
of decaying Burgers turbulence. 

\subsection{Velocity difference PDFs}

\subsubsection{Second moments: structure function}

\begin{figure}
\centering
\epsfig{file=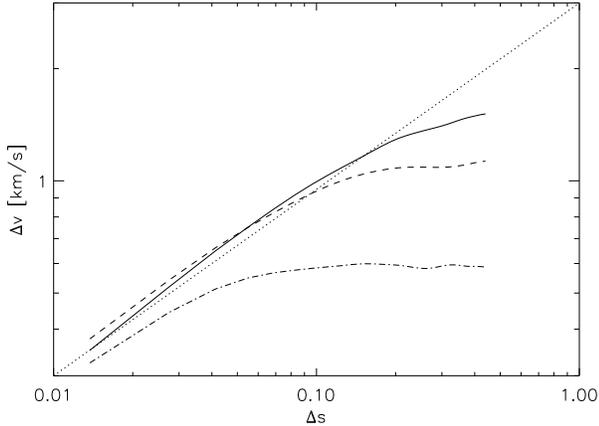,angle=90,width=\columnwidth}
\caption{Standard deviation of the centroid velocity difference PDF as a
function of lag, {\changed which is } equivalent to the structure function,
normalised to the square root of the lag for
3D velocity cubes of a hydrodynamic simulation driven
at wavenumbers of $k\sub{d}=2$ (solid), $k\sub{d}=4$ (dashed), and $k\sub{d}=8$
(dot-dashed).  These are models HE2, HE4, and HE8 from \cite{ml99}.}
\label{fig_newstructfn}
\end{figure}

In \S~\ref{sect_diffpdfs} we have shown that the second moment of 
the centroid velocity difference PDFs, or equivalently the structure 
function, is a differential measure for the size-linewidth relation,
so that the same power law was found for the two functions when applied
to the observations. In Fig.~\ref{fig_newstructfn} we plot the structure
function of the three models driven at different scales 
whose size-linewidth relation was shown in Fig.~\ref{fig_newlars}.

{\changed Comparison of the two figures demonstrates that the
structure function provides a better estimate of the driving
scale as its curvature is restricted to a narrower range 
around the driving scale than the curvature of the size-linewidth
relation. On the other hand, only the latter indicates the
dissipation scale by curvature at small lags. The different sensitivity 
of the two functions is caused by the
different shape of their effective filter functions 
(\S~\ref{sect_comparemethods}). }
We expect that with sufficient dynamic range there
would be a set of lags where both functions would show power-law
behaviour, but our $128^3$ models have such limited dynamic range that
there is effectively no {\changed scale} where this is true. 
The combined set of observations, on the other hand, does
have enough dynamic range to be dominated by the power-law slope, so
that both functions agree in most parts of the spectrum.

\subsubsection{Fourth moments: kurtosis}

The kurtosis of the centroid velocity difference distribution can
measure correlations in the motion at certain scales.  Values around
three indicate Gaussian distributions, implying motions uncorrelated
relative to the overall velocity field of the map considered.
\cite{Miesch99} found a decay of the kurtosis from small to large
scales roughly proportional to the square root of the scale down to a
value of three at the largest scales in their maps. In Sect.\
\ref{sect_diffpdfs} we found somewhat shallower slopes, and a flat
Gaussian section at scales above the total cloud size in the CfA
Polaris Flare map.

\begin{figure}
\centering
\epsfig{file=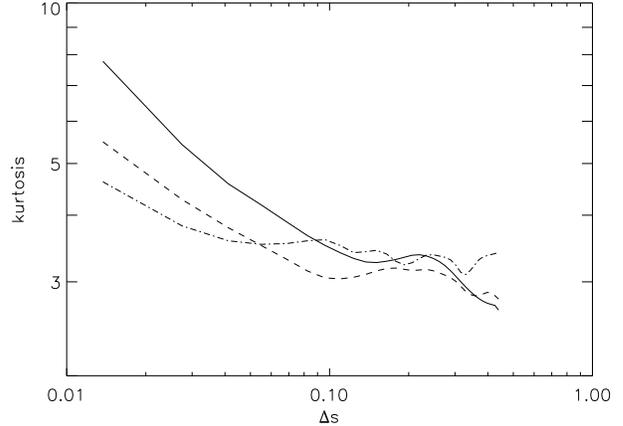,angle=90,width=\columnwidth}
\caption{Kurtosis of the centroid velocity difference distribution
for a hydrodynamic simulation driven at scales of $k\sub{d}=2$ (solid), $k\sub{d}=4$
(dashed), and $k\sub{d}=8$ (dash-dot).  These are 
models HE2, HE4, and HE8 from \cite{ml99}.}
\label{fig_kurtdriven}
\end{figure}
In Fig. \ref{fig_kurtdriven} we test how different driving scales in a
driven hydrodynamic model influence the resulting kurtosis plots. We
clearly see that the models driven at larger scales reach the Gaussian
value of kurtosis at larger lags, about a factor of two
below the peak size of the driving structure indicated by the
$\Delta$-variance. This suggests that motion at lags above the scale
of the driving process remains uncorrelated. At smaller lags, our
uniformly driven models show kurtosis following $\Delta
s^{-1/3}$, similar to the behaviour observed for the
Polaris Flare maps in Sect.\ \ref{sect_diffpdfs}. None of our models
reached the kurtosis values around 50
shown by the observations of Miesch et al.\ (1999), suggesting that
additional physics, especially in the driving function, may be
responsible for these high values.

Varying the resolution of our simulations does not appear to markedly
change the peak value of kurtosis, although the scale at which that
value is reached is always the smallest scale of the simulation,
suggesting that the correlated motions introduced by dissipation are a
major influence in producing non-Gaussian velocity-difference PDFs.
Magnetic fields do change the shape of the velocity
difference PDF, though not drastically. 
{\changed Increasing field strength increases the peak value of kurtosis 
by about 30\%.}

\subsection{$\Delta$-variance}

\label{sim_delta}
In paper~I we used the $\Delta$-variance to investigate the spatial
scaling of turbulent density structure.  Here, we use this method to
investigate the velocity structure.  In contrast to our analysis of
the observations, we have access to the full 3D structure in the
simulations.

\begin{figure}
\centering
\epsfig{file=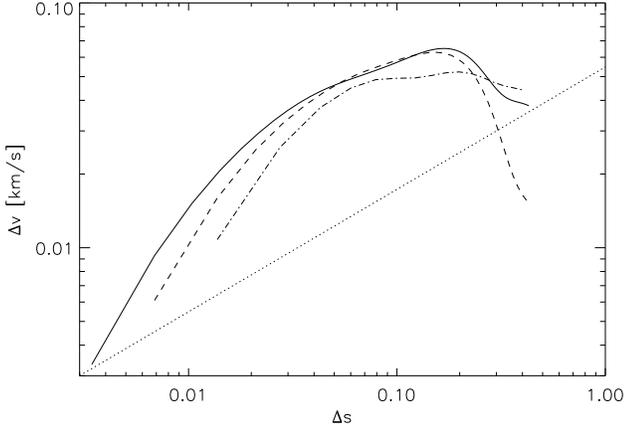,angle=90,width=\columnwidth}
\caption{$\Delta$-variance for
the velocity centroids for
3D velocity cubes of a hydrodynamic simulation of
decaying turbulence at time $t = 0.75 t_{\rm cross}$, computed at
resolutions of $64^3$ (dot-dashed), $128^3$ (dashed), and $256^3$ (solid). 
These are models B, C, and D from \cite{ml98}.}
\label{fig_newdeltares}
\end{figure}
We start by studying how changes in the resolution, and so in the
dissipation length scale, appear in the $\Delta$-variance spectrum.
In Figure~\ref{fig_newdeltares} we compare models of decaying
turbulence that are statistically identical, but were computed with
resolutions of $64^3$ (dash-dotted), $128^3$ (dashed), and $256^3$
(solid).  The variations at large scales ($\Delta s > 0.1$) are most
likely due to statistical fluctuations rather than the changes in
resolution. The result of systematically increasing the resolution can
be seen at small scales, as the spectrum reaches smaller scales at
higher resolution. The shape of the spectrum at the smallest scales in
each model is very similar, with only the scale changing.  This shows
the range over which numerical diffusion is acting.  Above that scale,
the models agree fairly well in a region that can be considered to be
the inertial range of the turbulence. We see that, equivalent to the density
structure, numerical diffusion causes a steepening of the spectrum,
and that this reaches to a scale of roughly ten zones in every model.

\begin{figure}
\centering
\epsfig{file=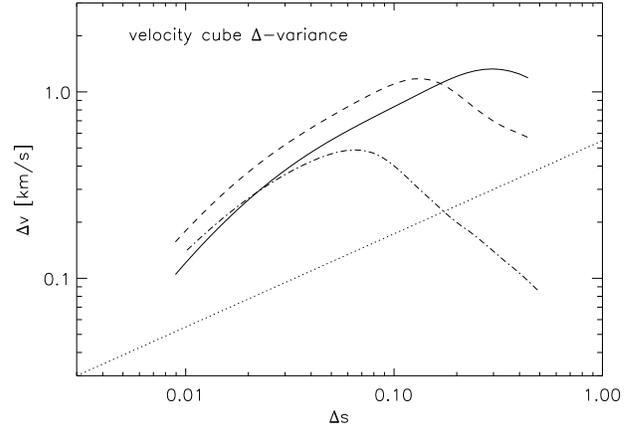,angle=90,width=\columnwidth}
\caption{$\Delta$-variances computed for 3D
velocity cubes of a hydrodynamic simulation driven with wavenumber $k\sub{d}
= 2$ (solid), $k\sub{d}=4$ (dashed), and $k\sub{d}=8$ (dot-dashed).  These are
models HE2, HE4, and HE8 from \cite{ml99}.}
\label{fig_delta3d}
\end{figure}
Now we can consider the effects of different driving wavelengths.
Fig.~\ref{fig_delta3d} shows the results of the analysis for three
hydrodynamic models driven at different scales (the $\Delta$-variance
for the density distribution in these models is shown in Fig.~7b of
paper~I). We plot the square root of the $\Delta$-variance, as that is
the quantity directly related to the linewidth. We see pronounced peaks close
to the driving scale at 0.5--0.6 $\lambda\sub{d}$.  They are even
somewhat more pronounced than in the density structure of these
models (paper~I), because the driving process itself is implemented in
velocity space (Mac Low 1999). Below the driving scale we find
power-law behaviour with a slope of 0.57 for $\Delta v$ corresponding
to a slope of 1.14 for the $\Delta$-variance, down to the scale of the
numerical viscosity at about ten pixels
($\Delta s \approx 0.04$ in the figure).

How does this slope compare to the expected power spectrum? As
discussed in paper~I, there is a theoretical relation between the
slope of the $\Delta$-variance $\alpha$ and the index $\zeta$ of an
$n$-dimensional power spectrum $P(\vec{k})\propto |\vec{k}|^{-\zeta}$
given by $\alpha=\zeta-n$.  Please note the difference to the index
$\zeta\sub{int}$ of the often used one-dimensional power spectrum
$P(k)$ that is obtained by angular integration of $P(\vec{k})$,
$\zeta=\zeta\sub{int}+n-1$. Hence, the slope is translated into values
of 4.14 and 2.14 for $\zeta$ and $\zeta\sub{int}$,
respectively. Incompressible Kolmogorov-type turbulence is
characterised by values of 11/3 and 5/3, respectively, while
shock-dominated velocity fields should show the Fourier transform of a 
step function, i.e. values of 4.0 and 2.0, respectively. 
The results from the simulations are quite close to the behaviour 
of shock-dominated gas, but the small deviation might ask for 
further investigation.

For the density structure, we found $\Delta$-variance slopes
between 0.45 and 0.75 in projection, corresponding to $\zeta$ 
({\em not} $\zeta\sub{int}$)
values between 2.45 and 2.75, considerably flatter than the slope in the
velocity structure found here. We speculate that this is due to the
compressibility of the isothermal gas modelled here.  The gas piles up
in thin sheets and filaments, so that the density structure has
behaviour somewhere in between that of a 
$\delta$ function and a step
function.  The Fourier transform of a one-dimensional $\delta$ 
function in the density cube is characterised by $\zeta=2$. The velocity on the
other hand, remains uniform for longer distances behind shocks, so
that its steeper spectral slope approaches more closely the value
expected for a box full of pure step functions.

\begin{figure}
\centering
\epsfig{file=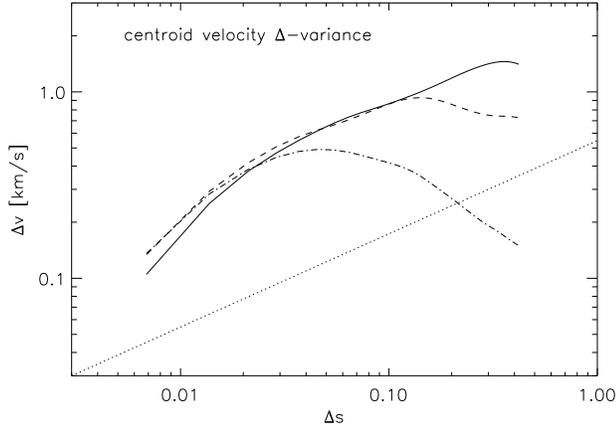,angle=90,width=\columnwidth}
\caption{$\Delta$-variances of the centroid velocity maps
of the three models from Fig.~\ref{fig_delta3d}, with driving
wavenumber $k\sub{d}=2$ (solid), $k\sub{d}=4$ (dashed), and $k\sub{d}=8$ (dot-dashed).}
\label{fig_deltacent}
\end{figure}
We can compare the 3D $\Delta$-variance spectrum to that 
from simulated velocity centroid maps. In Fig. \ref{fig_deltacent} we
show the $\Delta$-variance plots for the centroid velocity maps from
the models shown in Fig.~\ref{fig_delta3d}.  Here, identification of
the driving scale is no longer so easy.  Nevertheless, the general
functional behaviour is similar. The broad maximum is shifted
depending on the driving wavelength, and there is a power law range
with the slope close to 0.5. Thus, the $\Delta$-variance of the
velocity centroids does seem to reflect the true cloud velocity
structure, if not as clearly as the 3D spectrum. In the centroid
velocities we measure the same drift index $\alpha$ as in the 3D data
cube, in contrast to the density structure where the integrating
projection to two dimensions preserves the power spectral exponent
$\zeta$, and therefore increases the drift index $\alpha$ by 1
(\cite{Stutzki}). This conclusion also holds for the other models so
that we restrict ourselves here to the 3D $\Delta$-variance, noting
that the observational centroid maps do not reveal the structural
properties as clearly as discussed here.

{\changed Comparing Fig.\ \ref{fig_deltacent} with Figs.\
\ref{fig_newlars} and \ref{fig_newstructfn} or 
Fig.\ \ref{fig_newdeltares} with Fig.\ \ref{fig_newlarsres},
respectively, we see that the
$\Delta$-variance of the velocity centroids resolves both the driving
scale and the dissipation scale most clearly.  Although the structure
function also shows the driving scale by a change of the slope from
0.5 to 0, the bending from 0.5 to about -0.5 in the $\Delta$-variance
(Fig.~\ref{fig_deltacent}) can be detected more easily in noisy data.
The steepening at the dissipation scale which is also visible in
the size-linewidth relation by a change of the slope from 0.5 to 
about 0.8 is also greatest for the $\Delta$-variance where
a slope of about 1.2 is measured at the smallest scales.}

\begin{figure}
\centering \epsfig{file=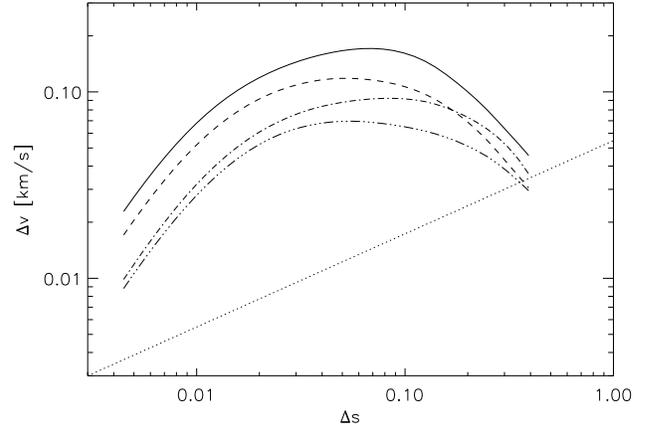,angle=90,width=\columnwidth}
\caption{$\Delta$-variance of the 3D velocity structure in a decaying
MHD turbulence model at times of 0.5, 0.75, 1.5, and 2.5 initial
crossing times.
\label{fig_simdecay}
}
\end{figure}
In Fig. \ref{fig_simdecay} we show the $\Delta$-variances for the 3D
velocity structure in a sequence of four time steps in the evolution of
strongly magnetised ($v\sub{A}/c\sub{s}=10$), decaying turbulence.
The magnitude of the variance drops over time, starting from small
scales, resulting in an increase in the effective peak of the driving
function and a slight increase in the effective slope of the spectrum
over time. This effect was also seen in the density structure in
paper~I, and there shown to be proportional to $t^{1/2}$ as predicted
by \cite{ml99}.

\begin{figure}
\centering
\epsfig{file=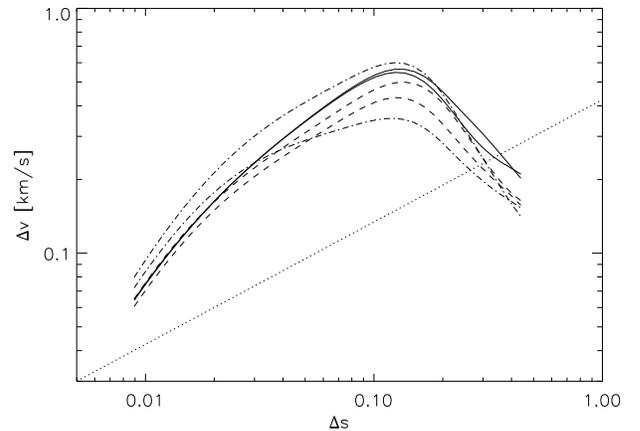,angle=90,width=\columnwidth}
\caption{$\Delta$-variance of the 3D velocity structure 
observed parallel (thick lines) and perpendicular (thin lines) to the
magnetic field for a models driven with $k\sub{d}=4$ and no field (solid),
weak field ($v\sub{A}/c\sub{s}=1$, dashed), and moderately strong field 
($v\sub{A}/c\sub{s}=5$, dash-dot).}
\label{fig_magnet}
\end{figure}
To demonstrate the effect of magnetic fields we show in
Fig.~\ref{fig_magnet} the 3D $\Delta$-variance plots observed parallel
and perpendicular to the initial magnetic field in two MHD simulations
with different field strengths, compared to a corresponding
hydrodynamic simulation.  We find a substantial
anisotropy in the case of a strong magnetic field {\changed (as in
\S~\ref{subsub-cenvelpdf})}, in contrast to the hydrodynamic
simulation and weak field cases. This is visible in the figure which
shows a velocity variance {\changed for the strong-field case (dash-dot 
curves)} substantially greater when viewed parallel to the initial field
direction (thick line) than perpendicular to it (thin line).

The theory of incompressible MHD turbulence (\cite{sg94,gs95,gs97})
predicts that the perpendicular cascade will be more efficient than
the parallel cascade, draining energy more quickly from perpendicular
motions and producing anisotropic wavevectors aligned parallel to the
field. Our results suggests that this continues to be
valid in the compressible regime.  Further support for this conclusion
may be drawn from the elongation of the density structures also seen
in the simulations (see Fig. 5 in Mac Low 1999).

In the weak field case (dashed line) the anisotropy is far smaller,
and indeed reversed, with perpendicular motions having slightly
greater power than parallel motions.  However, in this case, unlike
the strong field case, the field is weak enough that it is carried by
the flow into a tangled geometry, so that the initial field direction
no longer describes the geometry of the field well. The small
anisotropy observed is only a stochastic effect of the particular field 
configuration at that time. In fact, the plotted model is the only
example from our simulations where we find this reversal even at the
low field strengths.

Aside from the anisotropy, these driven models show little overall
difference in the velocity structure between a hydrodynamic model and
the corresponding MHD models. A small increase in the amount of
short-wavelength structure is seen with increasing magnetic field, as
in the density structure (paper~I), although there is no
substantial shift of the effective driving scale.  The small magnitude
of this effect supports the claim by Heitsch et al. (2000) that the
transfer of energy to smaller scales is insufficient to support
the smaller scales against gravitational collapse if they were able to
collapse in the absence of magnetic fields.

\section{Physical parameters}

\subsection{Observations}

The velocity structure in the Polaris Flare maps {\thirdchanged
shows basically a power-law behaviour below a few parsecs.
At the smallest scales we find some indication for a steepening
of the $\Delta$-variance spectrum of the velocity centroids,
but the noise level in the data prevents a definite conclusion here.
A similar scaling behaviour was demonstrated by \cite{Bensch}
for the intensity maps, reflecting the density structure. The
$\Delta$-variance spectrum in the intensity maps gradually
steepens from a slope $\zeta=2.6$ on large scales to $\zeta=3.3$ 
on the smallest scales. However, this behaviour is only visible
when combining results from multiple instruments with
different resolutions. The maps taken with each individual 
instrument appear to show only a power law within their limited
dynamic range.} The size-linewidth relation and structure function show
more constant slope, as is expected from their lesser sensitivity to
the driving scale.

The velocity centroid PDFs show nearly Gaussian wings, as do the 
average line profiles, though with different widths.  The PDFs of 
velocity difference as a function of lag, on the other hand, 
show non-Gaussian behaviour, with kurtosis values in some cases
exceeding the exponential value of six at the smallest scales,
indicating correlated motion on these scales. 

We discuss below the physical implications of these observations, as
deduced from our models.
 
\subsection{Model properties}

\label{model_properties}

The clearest result we can draw from the models is that
self-similar{\changed , power-law} behaviour in both velocity and
density structure can be found between the dissipation scale and the
driving scale even for the highly compressible turbulence that we
model here.  The behaviour is less clear but still definitely present
after projection into 2D.  The strength of the magnetic field also
does not strongly influence this conclusion, although stronger
magnetic fields modify the slope of the density spectrum and do
introduce measurable anisotropy into the turbulence, as shown in
Fig.~\ref{fig_magnet}.

Above the driving scale, a flattening or turnover of the spectrum is
apparent in all the measures we have studied, though it is most
pronounced in the $\Delta$-variance spectra.  Conversely, numerical
dissipation in the models causes a clear steepening of the spectrum as
structure at smaller scales disappears.  Although physical dissipation
will not have the same detailed properties as numerical dissipation,
the general steepening of the spectrum will certainly occur.  Similar
behaviour is observed in incompressible turbulence (Lesieur 1997).

The velocity centroid PDFs have kurtosis values in the range 2--4.
The lowest values are distinctly sub-Gaussian, and appear to be
produced only by driving at the largest available scales.  We agree
with \cite{k00} that these non-Gaussian PDFs may be due to
undersampling of a Gaussian distribution.  No other physics that we
have introduced produces such deviations.  Observations of
sub-Gaussian {\changed velocity centroid} PDFs thus is consistent with
driving from scales larger than the observed region.

The models driven at the largest scale show marked super-Gaussian
behaviour in their velocity difference PDFs at short lags, as reflected
in kurtosis values exceeding even the exponential value of six. 
The occurrence of such values is thus a strong indication 
that the primary driving occurs on scales as large as or larger than 
those in the map. 

\subsection{Turbulence properties}

Comparison between the observations of the Polaris Flare and the
turbulence simulations constrains the mechanisms driving the
turbulence in this cloud. Any mechanism that drives at an intermediate
length scale, such as jets from embedded protostars, should produce
characteristic features in the $\Delta$-variance at that scale, which 
are not observed. The approximately self-similar, power-law behaviour
seen in the observations is best reproduced by models where the energy
is injected at large scales and dissipated at the small scales.
{\thirdchanged The slight steepening of the $\Delta$-variance spectra
of the intensity structure, which seems also to be present in the 
velocity structure, is consistent with turbulence models where
the turbulence was recently injected and is decaying now due
to the dissipation (see paper~I).}

{\newchanged The large driving scale}
argues against protostellar outflows being the main driving mechanism.
An attractive alternative is driving by interactions with superbubbles
and field supernova remnants (e.g.\ {\changed Mac Low et al.\ 2001,}
Avillez et al. 2000, Norman \& Ferrara 1996).  The Polaris Flare
molecular cloud lies in the wall of a large cylindrical structure
representing one of the nearest H~{\sc i} supershells, the North
Celestial Polar Loop (Meyerdierks et al. 1991), adding additional
support to this proposal.

The kurtosis of the velocity difference PDFs in the Polaris Flare
observations, especially at small and intermediate scales, reaches
values close to 10, arguing for driving from scales larger than the
maps on which the high kurtosis appears, adding further support to the
interpretation given above.  We note that many of the actively
star-forming regions observed by Miesch et al.\ (1999) have even
higher kurtosis values at small scales, suggesting that the additional
physics of strong self-gravity and local heating may produce
additional effects that must be examined in future work.

{\newchanged The dominant physical mechanism} for dissipation in molecular 
clouds was first shown by \cite{zj83} to be ambipolar diffusion.
\mbox{}\cite{khm00} showed that the length scale on which ambipolar diffusion
will become important can be found by examining the ambipolar diffusion
Reynolds number
\begin{equation}
R_A = {\cal M}_A \tilde{L} \nu_{ni}/v_A
\end{equation}
defined by \cite{b96} and \cite{zb97}, where $\tilde{L}$ and ${\cal
M}_A $ are the characteristic length and Alfv\'en 
Mach number, $\nu_{ni} = \gamma \rho_i$ is the rate at which each 
neutral is hit by ions, and $v_A^2 = B^2/4\pi\rho$ approximates the 
effective Alfv\'en speed
in a mostly neutral region with total mass density $\rho = \rho_i +
\rho_n$ and magnetic field strength $B$.
The coupling constant
depends on the cross-section for ion-neutral interaction, and for
typical molecular cloud conditions has a value of $\gamma \approx 9.2
\times 10^{13}$~cm$^3$~s$^{-1}$~g$^{-1}$ (e.g.\ Smith \& Mac Low
1997).

Setting the ambipolar diffusion Reynolds number $R_A = 1$ yields a
diffusion length scale of
\begin{eqnarray}
L_D & = & v_A / {\cal M}_A \nu_{ni} \\
    & \approx & (0.041\mbox{pc}) {\cal M}_A 
      \left(\frac{B}{10\mbox{ $\mu$G}}\right)\!  
      \left(\frac{10^{-6}}{x}\right)\!
      \left(\frac{10^3 \mbox{ cm}^{-3}}{n\sub{n}}\right)^{3/2}
\end{eqnarray}
with the ionization fraction $x = \rho_i/\rho_n$ and the neutral
number density $n\sub{n} = \rho_n/\mu$, with $\mu = 2.36 m_H$.  If the
ionization level in the Polaris Flare is low enough {\changed and the
field is high enough}, this length scale {\changed of order 0.05 pc
would be} directly resolved in the IRAM observations. 
{\newchanged  We cannot yet unambiguously say whether the observed velocity
spectra show a steepening there, similar to the downturn at the dissipation
scale in the numerical models. If better observations do show such 
a downturn in the future, that will be an indication of the dissipation 
scale.}

\section{Summary}

We have applied several methods to characterise the velocity structure
observed in the Polaris Flare molecular cloud over scales ranging from
0.015~pc to about 20~pc.  We then applied the same methods to a
large suite of computational hydrodynamic and MHD models of
supersonic, isothermal turbulence.  The comparison between the
observations and models with different parameters allows us to draw
conclusions both about {\changed the properties of the analysis
methods and the physical state of the observed region:}
\begin{itemize}
\item By measuring the average variation of velocity centroids as a function
of the size of a virtual scanning beam in the observations and in
simulated observations, we recover Larson's (1981) size-linewidth
relation at scales where the turbulence shows an inertial range.  This
provides a method for measuring this relation that does not rely on
the identification of isolated clumps of gas. Comparing this variation
with the corresponding variation of the average total linewidth in
the virtual beams allows estimation of the depth of the cloud along
the line of sight.
\item The spectrum of the $\Delta$-variance, a multi-dimensional
wavelet transform (\cite{Stutzki}), shows
{\newchanged deviations from inertial
scaling behaviour at the scales of driving, dissipation, observational
noise, and the telescope beam} more clearly than {\changed
other methods that we tested}, 
{\changed because its effective
filter function is better confined in the spatial frequency domain}.
However the method currently lacks intensity weighting, so that it is 
not reliable on {\changed observed} maps with large empty areas.
\item {\changed  The structure function characterizing the
second moments of the velocity difference PDFs duplicates
information recovered by the scanning-beam
size-linewidth relation or the $\Delta$-variance of centroid velocities.
The fourth moments measure correlated motion within a map.}
\item We compare two proposed methods for measuring velocity PDFs by
comparing their results to the known PDFs in our models.  The average
line profile, determined in an optically thin line, is a better
measure than the distribution of line centroid velocities. The width
ratio of {\changed the two} distributions depends on the depth of the
observed region, as well as the Mach number of the flow, allowing us
to infer a Mach number exceeding 10 for the Polaris Flare. 
{\changed Sub-Gaussian velocity centroid PDFs are mainly produced by
the cosmic variance in the case of large-scale driving. The influence
of magnetic fields and driving strength appears weak.}
\item {\changed The power spectrum slope $\zeta$ can be derived 
from the slope of the $\Delta$-variance spectrum obtained for the
velocity centroid maps of the Polaris Flare.} {\newchanged At intermediate
scales in the KOSMA observations we find a slope in between 0.36 
and 0.46, corresponding to $\zeta = 3.7\dots 3.9$, while at small 
scales in the IRAM map we find $\zeta = 3.6\dots 4.4$.} 
{\thirdchanged Models of driven,
supersonic turbulence show inertial range slopes of $\zeta =
$3.9--4.2, with slopes steepening to $\zeta > 5$ in the dissipation
range.  (Dissipation is due to numerical viscosity in these models.)
The large error bars in the observed velocity spectra do not yet allow 
to draw a definite conclusion on a steepening in the dissipation 
range as seen in the models nor on a gradual steepening across the
full range of scales as seen by \cite{Bensch} in the intensity 
maps of the Polaris Flare. Future observations could do this.} 
\item The observations show super-Gaussian velocity difference PDFs at
small scales.  Only driving at the largest scales in our models
produces {\changed strongly super-Gaussian velocity difference}
PDFs at small lags.
\item The observed 
structure is consistent with hydrodynamic or MHD supersonic turbulence
showing a complete spectrum from a driving scale at larger than 10~pc,
through an inertial range, to a dissipation scale under
0.05~pc. {\changed The combination of the density and the velocity
scaling behaviour indicates a medium dominated by shocks at
intermediate length scales creating thin sheets or filaments.
}{\newchanged Ambipolar diffusion could explain the dissipation.}
\item {\changed The main uncertainties come from the noise in the
observational data leading to large error bars in the velocity scaling
functions and the unknown anisotropy of the velocity field.
Anisotropic velocity fields could be produced by large-scale driving
by supernovae or by strong magnetic fields but
cannot be detected in observations bound to the plane of the sky.}
\end{itemize}

\begin{acknowledgements}

We thank J. Ballesteros-Paredes, D. Balsara, F. Bensch, A. Burkert,
R. S. Klessen, M. D. Smith, J. Stutzki, \& E. Zweibel for useful
discussions, and the anonymous referee for helping us to clarify our
presentation. VO acknowledges support by the Deut\-sche
For\-schungs\-ge\-mein\-schaft through grants SFB 301C and 494B, and
M-MML acknowledges support by the NSF through a CAREER fellowship, AST
99-85392, {\changed and by the NASA Astrophysical Theory Program
through grant NAG5-10103.}  Computations presented here were performed
at the Rechen\-zentrum Garching of the Max-Planck-Gesell\-schaft, and
at the National Center for Supercomputing Applications (NCSA),
{\changed supported by the NSF}.  ZEUS was used by courtesy of the
Laboratory for Computational Astrophysics of the NCSA. This research
has made use of NASA's Astrophysics Data System Abstract Service.
\end{acknowledgements}


\begin{thebibliography}{}

\bibitem[Anselmet et al.\ 1984]{Anselmet}
Anselmet F., Gagne Y., Hopfinger E.J., \& Antonia R. 1984, J. Fluid Mech., 140, 63

\bibitem[Avillez et al. (2000)]{abm00}
Avillez M. A., Ballesteros-Paredes J., \& Mac Low M.-M. 2000,
Bull.\ Amer.\ Astron.\ Soc. 196, 26.08

\bibitem[Balsara (1996)]{b96}
Balsara D. 1996, \apj 465, 775

\bibitem[Bensch et al.\ (2001a)]{Bensch}
Bensch F., Stutzki J., \& Ossenkopf V. 2001a, \AandA 366, 636

\bibitem[Bensch et al.\ (2001b)]{Bensch2000}
Bensch F., Panis J.-F., Stutzki J., Heithausen A., \& Falgarone E. 2001b, 
\AandA 365, 275

\bibitem[Brunt (1999)]{Brunt}
Brunt Ch.\ M. 1999, {\em Turbulence in the Interstellar Medium},
PhD thesis, Univ. Massachusetts

\bibitem[Caselli \& Myers 1995]{caselli}
Caselli P. \& Myers P. C. 1995, \ApJ 446, 665

\bibitem[Chappell \& Scalo (1999)]{Chappell}
Chappell D. \& Scalo J. 1999, \MN 310, 1

\bibitem[Clarke 1994]{c94}
Clarke D. 1994, NCSA Technical Report

\bibitem[Dubinski et al. 1995]{Dubinski}
Dubinski J., Narayan R., \& Phillips T. G. 1995, \ApJ 448, 226

\bibitem[Evans \& Hawley 1988]{eh88}
Evans C. R. \& Hawley J. F. 1988, \apj 332, 659

\bibitem[Falgarone \& Phillips 1990]{Falgarone90}
{\newchanged Falgarone E. \& Phillips T.G. 1990, \ApJ 359, 344}

\bibitem[Falgarone et al.\ 1991]{Falgarone91}
Falgarone E., Phillips T. G., \& Walker C. K. 1991, \ApJ 378, 186

\bibitem[Falgarone et al.\ 1994]{Falgarone94}
Falgarone E., Lis D., Phillips T., et al. 1994, \ApJ 436, 728

\bibitem[Falgarone et al. 1995]{f95} 
Falgarone E., Pineau Des For\^ets G., \& Roueff E. 1995, \aap 300,
870 

\bibitem[Falgarone et al.\ 1998]{Falgarone}
Falgarone E., Panis J.-F., Heithausen A., et al. 1998, \AandA 331, 669

\bibitem[Goldreich \& Sridhar 1995]{gs95} Goldreich P. \& 
Sridhar S.\ 1995, \apj 438, 763 

\bibitem[Goldreich \& Sridhar 1997]{gs97} Goldreich P. \&
Sridhar S.\ 1997, \apj 485, 680 

\bibitem[Goodman et al. (1998)]{Goodman}
Goodman A. A., Barranco J. A., Wilner D. J., \& Heyer M. H. 1998,
\ApJ 504, 223

\bibitem[Hawley \& Stone 1995]{hs95} 
Hawley J. F. \& Stone J. M. 1995, Computer Phys. Comm. 89, 127

\bibitem[Heithausen \& Thaddeus (1990)]{Heithausen90}
Heithausen A. \& Thaddeus P. 1990, {\em ApJL} 353, L49

\bibitem[Heyer \& Schloerb (1997)]{Heyer}
Heyer M. H. \& Schloerb F. P. 1997, \ApJ 475, 173

\bibitem[Houlahan \& Scalo (1990)]{Houlahan}
Houlahan P. \& Scalo J. 1990, \ApJS 72, 133

\bibitem[Joulain et al.\ 1998]{j98} Joulain K., Falgarone E., Pineau des
For\^ets G., \& Flower D.\ 1998, \aap 340, 241

\bibitem[Kleiner \& Dickman (1985)]{Kleiner}
Kleiner S.C. \& Dickman R. L. 1985, \ApJ 295, 466

\bibitem[Klessen (2000)]{k00}
Klessen R. S. 2000, \apj 535, 869

\bibitem[Klessen et al. (2000)]{khm00}
Klessen R. S., Heitsch F., \& Mac Low M.-M. 2000, \ApJ 535, 887

\bibitem[Kolmogorov (1941)]{k41}
Kolmogorov A. N. 1941, {\em Dokl. Akad. Nauk SSSR} 30, 9

\bibitem[Larson 1981]{Larson}
Larson R. B. 1981, \MN 194, 809

\bibitem[van Leer 1977]{v77}
van Leer B. 1977, {\em J. Comput. Phys.} 23, 276

\bibitem[Lesieur (1997)]{l97}
Lesieur M. 1997, {\em Turbulence in Fluids}, 3rd ed., Kluwer
Dordrecht, 245

\bibitem[Lis et al. (1996)]{Lis}
Lis D. C., Pety J., Phillips T. G., \& Falgarone E. 1996, \ApJ 463, 623

\bibitem[Lis et al.\ (1998)]{Lis98}
Lis D. C., Keene J., Li Y., Phillips T. G., \& Pety J. 1998, \ApJ 504, 889

\bibitem[Mac Low (1999)]{ml99}
Mac Low M.-M. 1999, \ApJ 524, 169

\bibitem[Mac Low \& Ossenkopf (2000)]{paperI}
Mac Low M.-M. \& Ossenkopf V. 2000, \AandA 353, 339 (paper~I)

\bibitem[Mac Low et al.\ (1998)]{ml98}
Mac Low M.-M., Klessen R. S., Burkert A., \& Smith
M. D. 1998a, \prl 80, 2754

\bibitem[Mac Low et al.\ (2001)]{ml01}
{\changed Mac Low M.-M., Balsara D.S., Avillez M.A., \& Kim
J. 2001, \apj submitted} 

\bibitem[Meyerdierks et al. (1991)]{Meyerdierks}
Meyerdierks H., Heithausen A., \& Reif K. 1991, A\&A 245, 247

\bibitem[Miesch \& Bally (1994)]{Miesch94}
Miesch M. S. \& Bally J. 1994, \ApJ 429, 645

\bibitem[Miesch \& Scalo (1995)]{Miesch95}
Miesch M. S. \& Scalo J. M. 1995, \ApJ 450, L27

\bibitem[Miesch et al. (1999)]{Miesch99}
Miesch M. S., Scalo J., \& Bally J. 1999, \ApJ 524, 895

\bibitem[Myers 1983]{myers}
Myers P. C. 1983, \ApJ 270, 105

\bibitem[Norman \& Ferrara (1996)]{nf96}
Norman C. A. \& Ferrara A. 1996, \ApJ 467, 280

\bibitem{Ossenkopf99}
Ossenkopf V., Bensch F., Mac Low M.-M., \& Stutzki J. 1999, in Ossenkopf V., 
Stutzki J., \& Winnewisser G. (eds.), {\em The Physics and  Chemistry of the 
Interstellar Medium}, GCA-Verlag Herdecke, p. 216

\bibitem[Ossenkopf et al.\ (2000)]{Ossenkopf2000}
Ossenkopf V., Bensch F., \& Stutzki J. 2000, in: Gurzadyan V. G. \& Ruffini R.
(eds.), {\em The Chaotic Universe}, World Sci., p.394

\bibitem[Padoan et al.\ (1998)]{p98} 
Padoan, P., Juvela, M., Bally, J., \& Nordlund, {\AA}. 1998, \apj, 504, 300 

\bibitem[Padoan et al.\ (1999)]{p99} Padoan P., Bally J., 
Billawala Y., Juvela M., \& Nordlund {\AA}. 1999, \apj 525, 318 

\bibitem[Padoan et al.\ (2000)]{p00} 
Padoan P., Juvela M., Bally J., \& Nordlund {\AA}. 2000, \apj 529, 259 

\bibitem[Padoan \& Nordlund 1999]{pn99} Padoan P. \& 
Nordlund {\AA}. 1999, \apj 526, 279 

\bibitem[Peng et al.\ 1998]{peng}
Peng R., Langer W. D., Velusamy T., Kuiper T. G. H., \& Levin S.
1998, \ApJ 497, 842

\bibitem[Pety \& Falgarone 2000]{pf00} Pety, J. \& 
Falgarone {\'E}.\ 2000, \aap 356, 279 

\bibitem[Rosolowsky et al.\ 1999]{Rosolowsky}
Rosolowsky E. W., Goodman A. A., Wilner D. J., \& Williams J. P. 1999,
\ApJ 524, 887

\bibitem[She 1991]{She}
She Z.-S. 1991, {\em Fluid Dyn. Res.} 8, 143

\bibitem[Sridhar \& Goldreich 1994]{sg94} Sridhar S., 
Goldreich P. 1994, \apj 432, 612 

\bibitem[Stone \& Norman (1992a)]{sn92a} 
Stone J. M. \& Norman M. L. 1992a, \ApJS 80, 753

\bibitem[Stone \& Norman (1992b)]{sn92b}
Stone J. M. \& Norman M. L. 1992b, \ApJS 80, 791

\bibitem[Stutzki \& G\"usten (1990)]{sg90}
Stutzki J. \& G\"usten R. 1990, \apj 356, 513

\bibitem[Stutzki et al.\ (1988)]{s88}
Stutzki J., Stacey G. J., Genzel R., et al. 1988, \apj 332, 379

\bibitem[Stutzki et al.\ 1998]{Stutzki}
Stutzki J., Bensch F., Heithausen A., Ossenkopf V.,
\& Zielinsky M. 1998, \AandA 336, 697

\bibitem[Tauber (1996)]{Tauber}
Tauber J. A. 1996, \AandA 315, 591

\bibitem[V\'azquez-Semadeni (2000)]{Semadeni2000}
V\'azquez-Semadeni E. 2000, in Gurzadyan V. G. \& Ruffini R. (eds.), 
{\em The Chaotic Universe}, World Sci., p.384 

\bibitem[V\'azquez-Semadeni et al.\ (1997)]{Semadeni}
V\'azquez-Semadeni E., Ballesteros-Paredes J., \& Rodriguez L. F. 1997,
\ApJ 474, 292

\bibitem[Zagury et al. (1999)]{Zagury}
Zagury F., Boulanger F., \& Banchet V. 1999, A\&A 352, 645

\bibitem[Zweibel \& Brandenburg (1997)]{zb97}
Zweibel E. G. \& Brandenburg A. 1997, \ApJ 478, 563 (err: 485,
920)

\bibitem[Zweibel \& Josafatsson (1983)]{zj83}
Zweibel E. G. \& Josafatsson K. 1983, \apj 270, 511

\end{thebibliography}
\end{document}